# Hall effect and symmetry breaking in non-magnetic metal with dynamic charge stripes

N. Sluchanko[1], A. Azarevich[1], A. Bogach[1], S. Demishev[1], K. Krasikov[1], V. Voronov[1],

V. Filipov[2], N. Shitsevalova[2], V. Glushkov[1]

[1] A.M. Prokhorov General Physics Institute of the Russian Academy of Sciences, 38 Vavilov Str., 119991 Moscow, Russia

[2] Frantsevich Institute for Problems of Materials Science, National Academy of Sciences of Ukraine, 3 Krzhyzhanovsky Str., 03680 Kyiv, Ukraine

A comprehensive study of magnetoresistance and Hall effect has been performed for the set of the single crystals of non-magnetic metal $LuB_{12}$ with the Jahn-Teller instability of the boron cage and dynamic charge stripes forming along <110> direction. Anomalous positive contribution to Hall effect for particular direction of magnetic field **H**//[001] is found in the single crystals of $LuB_{12}$ of the highest quality. This contribution arising at T~ 150 K is shown to increase drastically when approaching the disordered ground state below 60 K. The Hall effect anomaly is shown to appear in combination with the peak of magnetoresistance. The various scenarios allowing for the topology of Fermi surface, anisotropy of relaxation time for charge carriers and the filamentary structure of fluctuating charge stripes are proposed to explain the features of magnetotransport in this metal with inhomogeneous distribution of electron density. The origin of SdH oscillations, which are observed in this non-equilibrium metal with electron phase separation and strong charge carrier scattering, is discussed.

**1. Introduction.** The unique opportunity for studying of the features of electron phase separation in strongly correlated electron systems is provided by dynamic, or fluctuating charge and spin stripes, which attract extraordinary scientific interest due to their important role in the mechanism of high temperature superconductivity (HTSC, see, e.g., [1-14]). While the stripe states are generally considered as a special feature for the $La_{2-x}Ba_xCuO_4$ perovskite family [15], phase singularities accompanied by lowered symmetry of the lattice are detected in other cuprates ($YBa_2Cu_3O_{7-\delta}$ [16-17], $SmBa_2Cu_3O_x$ [18], $Bi_2Sr_2CaCu_2O_{8+\delta}$ [19] and $Ca_{2-x}Na_xCuO_2Cl_2$ [20]). This kind of electron instability is also suggested to be extremely important for understanding of exotic physics in colossal magnetoresistive manganites [21-22], nickelates [23-25], iron-based superconductors [26-27], heavy fermion hexaborides [28], etc.

An especial kind of fluctuating charge stripes have been recently detected in the non-magnetic dodecaboride $LuB_{12}$ [29,30,31] and magnetic $Tm_{1-x}Yb_xB_{12}$ compounds with metal-insulator transition [32]. The mechanism responsible for this electron and lattice instability in $RB_{12}$ with *fcc* crystal structure is related to dynamic cooperative Jahn-Teller (JT) effect in the $B_{12}$ clusters [29,31]. It was shown in [29] that, because of triple orbital degeneracy of the ground electronic state, the $B_{12}$ molecules are JT active and their structure is labile due to JT distortions. The quantum chemical calculations and geometry optimizations for the charged $[B_{12}]^{2-}$ cluster, which doubly negative charge state is regarded as the most relevant one in $RB_{12}$ compounds [29], allow concluding in favor of strong trigonal and tetragonal distortions of the *fcc* lattice and mixture via a non-adiabatic electron-vibronic coupling of electronic states for each $B_{12}$ cluster [33]. In the dodecaboride matrix (see fragment of the *fcc* crystal structure in Fig.1a), the *collective JT effect* in the lattice of these $B_{12}$ complexes is the origin of both the collective dynamics of boron clusters and large amplitude vibrations (rattling modes) of the rare earth ions embedded in $B_{24}$ cavities of the boron cage (see Fig.1b). Strong coupling of these Lu rattling modes located at 110 cm$^{-1}$ (~14 meV) [34] and the JT active vibrations of boron are the origin of both lattice instability and the emergence of the collective excitations in the optical spectra of LuB$_{12}$ [35].

The Lu rattling modes correspond to Einstein oscillators in the double-well potential (Fig.1c). The large amplitude of these rattling modes initiates strong changes in the *hybridization* of rare earth and boron electron states. Since the states in the conduction band are contributed by the *2p* orbitals of B$_{12}$ and the *5d* orbitals of Lu [36, 37, 38, 39], the variation of the *5d–2p* hybridization leads to the modulation of conduction band width and, consequently, may generate non-equilibrium ("hot") electron states. These charge carriers seem to amount by ~70% from the total number of conduction electrons as estimated from overdamped oscillators in the optical conductivity spectra at room temperature [35]. Due to the stripe formation a considerable part of non-equilibrium conduction electrons is accumulated in the filamentary structure of fluctuating charges (see wide green channels in Fig.1a [30]). Note also that the lattice and electron systems of LuB$_{12}$ are significantly modified below $T^*$ ~ 60 K where the cage-glass state [40] is formed due to positional disorder of the loosely bound Lu$^{3+}$ ions embedded in B$_{24}$ truncated cuboctahedrons caused by static displacements of R$^{3+}$ ions in the double-well potential (Fig.1b-c).

The investigation of crystal structure, heat capacity and charge transport [31] allowed to argue that in the family of Lu$^{N}$B$_{12}$ crystals with various boron isotopes N=10, 11, nat (the last index corresponds to the natural mixture of 18.8% $^{10}$B and 81.2% $^{11}$B) Lu$^{nat}$B$_{12}$ possesses the strongest local atomic disordering in combination with long-range JT trigonal distortions. These factors provide optimal conditions for the formation of the dynamic charge stripes below $T^*$~60 K. As a result, the resistivity and Seebeck coefficient of the Lu$^{nat}$B$_{12}$ heterogeneous medium decrease strongly in comparison with those ones detected for pure boron isotope crystals. It was concluded in [31] that any defects are supposedly used as the pinning centers facilitating the formation of additional quantum conductive channels (dynamic charge stripes) in the metallic matrix of RB$_{12}$.

Hall effect is usually considered as a key experiment to reveal a transformation of electron spectrum in the vicinity of quantum critical point (QCP) [41-44], in the stripe-ordered compounds [45-46], near the metal-insulator transition [47-49], etc. All the singularities were detected in the $Tm_{1-x}Yb_xB_{12}$ family [50, 32] and the complicated behavior of the Hall resistivity components was interpreted in terms of both the metal-insulator transition and electron phase separation effects in combination with formation of nanoclusters of the rare earth ions in the cage-glass state of the studied compounds. The instability of Yb 4$f$-electron configuration in these Yb-based rare-earth dodecaborides is developed in combination with the JT deformations of the boron cage (Yb ion valence is about 2.92-2.95 [51-52]). This phenomenon provides one more mechanism of the charge and spin fluctuations in $RB_{12}$ and it is promising to separate only the influence of dynamic charge stripes on the Hall voltage anomalies in $RB_{12}$. In this study we present the results of detailed Hall effect measurements of different quality single crystals of the non-magnetic reference compound $LuB_{12}$. We show here that an ordinary Hall effect in $LuB_{12}$ is modified by the appearance of specific contribution to the Hall resistivity, which could not be explained exclusively in terms of anisotropic Fermi surface and is suggested to be caused by an interaction of the dynamic charge stripes with external magnetic field.

## 2. Experimental details.

**2.1. Crystal growth.** The single crystals of non-magnetic $LuB_{12}$ dodecaboride were grown using the induction crucible-free zone melting method with multiple re-melting in an argon gas atmosphere from the preliminarily synthesized $LuB_{12}$ powders [53]. All the crystals under investigation (#1-6) have been grown from the initial high purity $LuB_{12}$ powder synthesized from the 4N lutetium oxide $Lu_2O_3$ (99.9985 wt. % purity, particle size less than 100 microns) and 3N amorphous boron (99.9 wt. % purity, particle size about 50 Å). The details of both ingot preparation and solid state reaction are given in [54]. Different technological regimes were applied during the growing process. In the case of crystal # 1, an additional purification of argon from possible contamination of other gases was carried out directly in the growth chamber where Ti was used as a getter. The crystals #1-2 and #4-6 were grown with single zone passing under different pressures of argon gas ranging from 0.15 to 0.5 MPa with equal crystallization rates. The crystal #3 with the highest residual resistivity value was grown with triple zone passing. Perfect x-ray diffraction reflexes in Laue pattern (for crystals #1-6) and quantum oscillations of the magnetization (de Haas-van Alphen effect) observed in magnetic field **H**‖[100] (for crystals #1) prove the high quality of the crystals under investigation.

### 2.2. Galvanomagnetic measurements.

The samples for magnetoresistance (MR) and Hall effect (HE) measurements were cut from the single domain crystals oriented along the principal crystallographic directions with accuracy ~2º, then these were polished and etched to delete the distorted surface layer. The angular dependences of resistivity and Hall voltage were measured by the original sample rotation technique with a stepwise ($\Delta\varphi=1.8°$) fixing of the sample position in the steady magnetic field. The magnetic field supplied by superconducting solenoid up to 80 kOe was applied perpendicular to the measuring direct current (DC) **I** ‖ <110> in #1-#6 (see inset in Fig. 5a below, the orientation of the normal vector **n** ‖ <001>, or **n** ‖ <110> was used). For the samples #1 the resistivity and Hall effect measurements were carried out also in the magnetic field up to 140 kOe at liquid helium temperature (sample 1a, **I** ‖ [110]) and for #1 and #2 in various DC directions (samples #1b and #1c with **I** ‖ [100] and **I** ‖ [010], correspondingly). The installation equipped with a step-motor with automated control of the step-by-step sample rotation is similar to that one applied previously in [50]. High accuracy of the temperature control ($\Delta T \approx 0.002K$ in the range 1.8-7K) and magnetic field stabilization ($\Delta H \approx 2$ Oe) was achieved by using the commercial temperature controller TC 1.5∕300 and superconducting magnet power supply unit SMPS 100 (Cryotel Ltd.) in combination with CERNOX 1050 thermometer (Lake Shore Cryotronics, Inc.) and Hall sensors.

### 3. Experimental results and analysis.

**3.1. Temperature dependences (I//<110>).** Fig. 2 shows the metallic type temperature dependences of resistivity $\rho(T)$ measured for various $LuB_{12}$ single crystals. The temperature independent "plateau" is observed on the $\rho(T)$ curves in the range $T \leq 20$ K for all the studied crystals. The residual resistivity ratio (RRR=$\rho$(300 K)/$\rho$(4.2 K)) varies about by the factor of 6 from ~70 ( sample #1) to ~12 (sample #3). The resistivity $\rho(T, H_0)$ and the reduced Hall resistivity $\rho_H/H = f(T, H_0)$ have been studied in detail in magnetic field $H_0$=80 kOe for samples #1, #2 and #3 in two orientations **H**//[100] and **H**//[110] (see Fig.3). Taking into account the electron density inhomogeneity in $LuB_{12}$ due to dynamic charge stripes found in the dodecaboride matrix [30] (see Fig. 1a), the term "reduced Hall resistivity" $\rho_H/H = f(T, H_0)$ will be used below instead of Hall coefficient $R_H(T, H_0) \equiv \rho_{xy}/H$. Fig. 3 presents the data of Hall resistivity measured for two opposite directions of external magnetic field $\pm$**H**||**n** and calculated applying the general relation $\rho_H = [(V_H(H) - V_H (-H))/( 2I)]d$ ($V_H$ is the voltage from the Hall probes and $d$ is the thickness of the sample or, equivalently, its size along the normal direction to the largest face). For the sample #1a strong MR anisotropy is clearly seen in Fig.3a with a maximum values detected for **H**//[100] and it is accompanied with a dramatic decrease of the absolute values of $\rho_H/H$ with the temperature lowering (Fig.3b). On the contrary, the reduced Hall resistivity $\rho_H/H$ for **H**//[110] reveals only moderate changes with a negative minimum at the cage-glass transition temperature $T^* \sim 60$ K [40]. Thus, the values of $\rho_H/H$ in crystal #1a differ by more than 3 times for these two directions of magnetic field. Simultaneously both the MR anisotropy and the strong temperature variation of the reduced Hall resistivity $\rho_H/H$ for **H**//[100] depress essentially with the increase of residual resistivity $\rho_0$ in the sample #2 (Fig. 3 c,d). This low temperature anisotropy becomes very small in the sample #3 (Fig.3 e,f) with the highest $\rho_0$ value (shortest mean free path of the charge carriers). Taking into account that the Hall coefficient of normal uncompensated metal in strong magnetic fields is determined by the difference of electron and hole concentrations of $R_H = \rho_H/H \sim (n_e - n_h)^{-1}$ [55,56] it is extremely unusual to observe such a strong anisotropy of the reduced Hall resistivity $\rho_H$/H for the highest quality crystal (#1a) in the $LuB_{12}$ family (#1-#6).

Fig. 2b demonstrates the temperature dependences of Hall mobility $\mu_H(T)= (\rho_H(T)/H)/ \rho(T)$ estimated for the crystals #1, #2 and #3 for two directions **H**//[100] and **H**//[110]. It is worth noting that unusually strong difference between $\mu_H$(**H**//[100]) and $\mu_H$(**H**//[110]) (about 6 times) appears for the sample #1a just below $T^*$~60 K in the disordered cage-glass phase of $LuB_{12}$. The ratio $\mu_H$(**H**//[100])/$\mu_H$(**H**//[110]) is reduced by a factor of 3 for sample #2 and becomes as low as 1.1 in the case of the sample #3. The exponential behavior of Hall mobility $\mu_H(T)$~$T^\alpha$ is observed in the temperature range 80-300 K for all $LuB_{12}$ crystals; the $\alpha\approx$7/4 exponent is detected for the samples #1 and #2 with the large values of RRR=40-70 and $\alpha\approx$3/2 is estimated for the sample #3 (RRR~12) (Fig.2b). Note that the $\alpha$=3/2 exponent is typical for the scattering of conduction electrons by acoustic phonons (deformation potential) and the increase of $\alpha$ values up to 7/4 in the crystals #1 and #2 means that other scattering channels contribute mainly to the relaxation processes of charge carriers in the clean limit of $LuB_{12}$. The details of the phonon spectra of $LuB_{12}$ have been detected in [34] allowing to conclude in favor of the dominant carriers scattering on the quasi-local vibrations of the RE ions. So, relaxation processes due to scattering on both the boron optical phonons [34] in the presence of the JT distortions and the rattling modes of the RE ions [29-31] may be considered as dominant factors, which are responsible for the observed $\alpha$ increase in the samples #1 and #2 of the RE dodecaboride with JT instability of the boron sub-lattice.

**3.2. Magnetic field sweeping (I//<110>).** The magnetic field dependences of reduced Hall resistivity $\rho_H$/H=f(H) (Fig.4a), resistivity $\rho(H)$ (Fig.4b) and Hall mobility $\mu_H(H)$ (Fig.4c) show the results obtained for the sample #1a at T=4.2 K (see also Supplementary Information [54] for other $LuB_{12}$ crystals). For **H**//[100] the amplitude of the reduced Hall resistivity demonstrates a dramatic (~10 times) decrease with the rise of magnetic field up to 140 kOe, and, on the contrary, for **H**//[110] the amplitude of $\rho_H/H=f(H)$ increases only slightly. It is worth noting that the low-field values of $\rho_H$/H for **H**//[110] and **H**//[100] are just equal to each other (Fig.4a), so any noticeable discrepancy is developed only above 10 kOe. Straightforward estimations of Hall mobility for the highest quality crystal #1a from the data of Figs.4a and 4b result in high enough values of $\mu_H$ ~3000 cm$^2$/(V s) in low fields and small values of $\mu_H(H\approx140$ kOe)~18 cm$^2$/(V s) in high fields for **H**//[100] at liquid helium temperature. Note, that the last values are about 40 times smaller than those ones detected for **H**//[110] (Fig.4c). This observation proves that simple free electron model fails to extract any real miscroscopic parameters of charge carriers from the experimental data. Taking into account values of Fermi velocity $v_F\approx1.1\cdot10^7$ cm/s [35] and the effective mass $m^*$~0.5 $m_0$ [38] small enough average mean free path of charge carriers $l$~350 Å may be estimated even in the best quality single crystal #1 of LuB$_{12}$. It is worth noting also that the difference between estimated Hall mobilities for **H**//[100] and **H**//[110] decreases strongly for the sample #2 with $\mu_H(H<10$ kOe)~1500 cm$^2$/(V s) and becomes much smaller for the sample #3 where $\mu_H(H<10$ kOe)~420 cm$^2$/(V s) (see [54] for more detail) giving evidence for strong degradation of mean free path in the #1-#2-#3 series.

**3.3. Transverse MR anisotropy (I//<110>).** The anisotropy of the transverse MR $\Delta\rho/\rho$ in LuB$_{12}$ at liquid helium temperature is analyzed by discussing of (*i*) the magnetic field dependences of magnetoresistance collected at different fixed angles φ between the normal **n** to the (100) surface and the applied magnetic field ($\angle\varphi=(\mathbf{n}^{\wedge}\mathbf{H})$, see the inset of Fig. 5a) and (*ii*) the angular dependence of magnetoresistance measured in various magnetic fields up to 80 kOe. The data of field and angle sweeping measurements for the sample #1a are shown in Figs. 5a and 5b, respectively. For comparison, Fig. 5c shows the anisotropy of MR detected for several LuB$_{12}$ crystals having different RRR values (from 12 to 70) and the liquid helium Hall mobility ranging as 420-3000 cm$^2$/(V s) (see Fig.4 and the data in [54]). It is easy to see that a whole set of extrema with small enough values of MR is detected on the $\Delta\rho/\rho(\varphi)$ dependences in the angle range 54°<φ<126° in the vicinity of the diagonals **H**//[110] and **H**//[111]. At the same time, $\Delta\rho/\rho$ strongly increases within a wide range of angles ($-25°<\varphi<+25°$) when approaching to **H**//[001] (Fig. 5b and 5c). Note that the observed anisotropy of transverse magnetoresistance $\Delta\rho/\rho(\varphi)$ is very similar to this one reported earlier for $LuB_{12}$ crystal at 0.5 K in 12 T [38]. This distorted cross-type anisotropy of MR is mostly pronounced in the polar plot shown in Fig. 6a and it could be evidenced either in favor of the existence of an open trajectories on the Fermi surface [38], or, it shows up the presence of the filamentary structure of the dynamic charge stripes in $LuB_{12}$.

To shed more light on the details of the MR anisotropy in the non-magnetic compound $LuB_{12}$ the angular dependences of $\Delta\rho/\rho(\varphi, T_0, H_0)$ have been studied for the sample #1a in the strong magnetic field $H_0$=80 kOe in the wide temperature range 1.8-200 K. Fig. 6b shows the 3D view of the transverse MR in the cylindrical plot. It is easy to see in Fig.6b that the strongest increase of MR occurs when **H** is oriented perpendicular to the filamentary structure of the quantum conduction channels (**H**//[001]) while only moderate humps of MR are detected in the wide neighborhood of **H**//[110] dynamic charge stripe direction. Interestingly, both of the mentioned features are developed just below the cage-glass transition $T^*\sim60$ K (marked by dotted white line in Fig.6b).

### 3.4. Angular dependences of Hall resistivity (**I**//<110>).

The selected angular dependences of reduced Hall resistivity $\rho_H/H$=$f(\varphi, H_0)$ measured in various magnetic fields at liquid helium temperature are presented for two different orientations of the normal vector **n**//[001] and **n**//[110] in Figs. 7 and 8, correspondingly (the sample #1a, current axis **I**//[110]; see also fig.S2 in [54] for the sample #2). The cosine-like behavior of Hall resistivity angular dependences is natural for isotropic non-magnetic metal reflecting the changes of potential difference between two Hall probes due to varying transverse HE electric field (see the schema in the inset in fig.5a). This simplified behavior is really observed for the sample #3 (Figs. 7c, 8c) with smallest RRR, where Hall mobility is less than 450 $cm^2$/(V s) (Fig. S3 in [54]). On the contrary, an additional anomalous component contributes to Hall signal for the samples #1a and

#2 in the configuration $\mathbf{I}//[110]$, $\mathbf{n}//[001]$ in the wide neighborhood of $\mathbf{H}//<001>$ direction (Fig.7). The singularity is strongest for the sample #1a with the largest RRR ($\mu_H$(10 kOe)~3000 cm$^2$/(V s), Fig.4c). Indeed, the magnitude of the anomalous non-harmonic contribution for the sample #1a at H=80 kOe compensates substantially of the cosine-like component of the ordinary Hall effect near $\mathbf{H}//\mathbf{n}//[001]$, and the non-harmonic feature depresses for the samples #2 and #3 (see Figs.7a, 7b and 7c, and Fig.S6 in [54]). Moreover, the singularity is not observed when $\mathbf{H}//[110]$ for the same rotation axis ($\mathbf{I}//[1-10]$) and the other $\mathbf{n}//[110]$ configuration). Instead, two additional anomalous features of opposite sign have been detected on the $\rho_H/H=f(\varphi, H_0)$ data in the vicinity of $\mathbf{H}//<111>$ directions (see Fig.8 for #1a).

The dramatic difference between $\rho_H/H$ vs. T (Fig.3) and $\rho_H/H$ vs. H (Fig.4) for $\mathbf{H}//[001]$ and $\mathbf{H}//[110]$ detected in our experiments and the qualitative dissimilarity of the $\rho_H/H=f(\varphi, H_0)$ data for $\mathbf{H}//[001]$ and $\mathbf{H}//[110]$ (shown by arrows in Figs.7a and 8a, respectively) make evident that the discovered singularity in the vicinity of $\mathbf{H}//<001>$ is responsible for very strong changes in the Hall signal observed in the $\mathbf{H}//[001]$ configuration. Indeed, the amplitude of the feature detected near $\mathbf{H}//<001>$ increases drastically both in magnetic field (see Figs. 7a, 9b) and with temperature decreasing below $T^*$~60 K (see Fig. 9a). It should be pointed out that the anomaly is observed in the sample #1a (see also Fig.S7 in [54] for the sample #2) when the external magnetic field $\mathbf{H}//<001>$ is applied perpendicular to the [110] direction of dynamic charge stripes (see Fig. 1a). This observation points to the possible relationship between the observed singularity and the effects of interaction of magnetic field with the filamentary structure of the quantum conduction channels. It is worth noting also, that in low magnetic field (H<20 kOe) the amplitude of the cosine-like contribution decreases essentially (~15%) in the sample sequence #1- #2- #3 (fig.7), so, it should be suggested that the emergence of the infinite cluster of the dynamic charge stripes is accompanied with the moderate enhancement of the negative ordinary HE. More detailed quantitative analysis and general discussion of HE in LuB$_{12}$ is presented below.

**3.5. H-T variation of HE components (I//<110>).** To separate the ordinary and anomalous contributions to Hall effect the experimental data $\rho_H(\varphi, H_0, T_0)$ (Figs. 9a-b and 10) were fitted by cosine law $\rho_{H0}(H_0,T_0)\cos\varphi$ in the segments $90º\pm20º$ and $270º\pm20º$ (see, *e.g.*, Fig.9b and 10 for #1a and Fig.S7 in [54] for #2). The difference $\rho_H^{an}(\varphi, H_0, T_0)=\rho_H(\varphi, H_0, T_0)-\rho_{H0}(H_0,T_0)\cos\varphi$ was attributed to the additional specific component of HE. The temperature and field dependences of the main term $R_H(T, H) \equiv \rho_{H0}/H$ (ordinary Hall effect, marked as "cos") are shown in Figs. 3b and 4a for the sample #1a, and in Figs. 3d and in [54] for the sample #2. The H-T variation of the additional term $\rho_H^{an}/H(\varphi, H_0, T_0)$ for **n**//[001] configuration is presented in Figs. 9c and 9d for #1a and in Fig.S7 in [54] for #2. Fig. 10 shows the reduced Hall resistivities $\rho_H/H$ and $\rho_H^{an}/H=f(\varphi, H_0, T_0)$ detected in the **n**//[110] configuration of #1a at 4.2K. It is clearly discerned from Figs.9c, 9d that the largest $\rho_H^{an}/H$ is observed for the **n**//[001] configuration around the <001> axis and the positive polarity of the additional specific component is opposite to the negative ordinary contribution in the Hall voltage of LuB$_{12}$. Besides, this anomalous term consists of two parts − smooth $\Delta_{smooth}(\varphi, H_0, T_0)$ and step-like $\Delta_{step}(\varphi, H_0, T_0)$ (see the legend in Figs. 9c, 9d). The data analysis allows detecting the changes in these $\Delta_{smooth}$ and $\Delta_{step}$ amplitudes both with the temperature variation (Fig. 9e) and magnetic field increasing (Fig.9f). Note that the sum of the specific component and ordinary Hall coefficient within the experimental accuracy is equal to Hall resistivity $\rho_H/H=f(\varphi, H_0, T_0)$ measured without sample rotation as a difference between the data for two opposite directions of magnetic field (see Figs. 3a and 4a), arguing in favor of the proposed explanation of the Hall voltage anomalies in the metal with dynamic charge stripes.

It is easy to see from Figs.9e, 9f that the main specific component $\Delta_{smooth}$ appears in the crystals #1 and #2 of LuB$_{12}$ just below $T_E\sim150$ K, where the development of the JT instability of B$_{12}$ sub-lattice was found previously in various experiments. Indeed, it was shown in Raman studies [40] that temperature lowering leads to a sharp increase in the vibration density of states near $T_E$ when the mean free path of phonons reaches the Ioffe–Regel limit and becomes comparable with their wavelength [57]. Moreover, the sharp peak of the relaxation rate in the vicinity of $T_E$ is observed in $\mu SR$ experiments with $RB_{12}$ dodecaborides (R = Er, Yb, Lu) and $Lu_{1-x}Yb_xB_{12}$ solid solutions [58,59]. The authors [58, 59] suggested that the large amplitude dynamic features in the $\mu SR$ spectra are induced by the atomic motions within the $B_{12}$ clusters.

Below $T^*\sim 60$ K in the cage-glass state of $LuB_{12}$ [40] the step-like $\Delta_{step}$ component appears additionally to the $\Delta_{smooth}$ contribution (see Fig. 9c, 9e) and the amplitudes of these two terms increase simultaneously with the temperature lowering. At T=4.2K these anomalous positive components are not observed in low magnetic fields H≤10 kOe, and the only $\Delta_{smooth}$ contribution is detected on the angular dependences of the SIHE in the range 10-30 kOe (Figs. 9d, 9f). Then, $\Delta_{step}$ feature appears above $H^*$=30 kOe, and the $\Delta_{step}(H)$ and $\Delta_{smooth}(H)$ dependences are found to be quite different. Indeed, the smooth component saturates in the fields up to 80 kOe, while the step-like one increases strongly resulting to linear increase of the total amplitude (Fig. 9f). When RRR decreases in the studied $LuB_{12}$ crystals the additional $\Delta_{smooth}$ contribution is only observed with the linear field dependence (see curve for the sample #2 in Fig.9f).

The analysis of the temperature dependence $\rho_H^{an}/H=f(T, H_0=80$ kOe) in the crystals #1 and #2 allows us to reveal a hyperbolic type behavior $\rho_H^{an}/H=C^*(1/T-1/T_0)$ (Fig. 9e). The similar trend has been previously detected for anomalous transverse even component of Hall voltage in strongly correlated electron systems (SCES) $CeCu_{6-x}Au_x$ [60] and $Tm_{1-x}Yb_xB_{12}$ [50]. Similar to the results for $LuB_{12}$ (Fig. 9e) the anomalous hyperbolic component in $Tm_{1-x}Yb_xB_{12}$ appears in high enough magnetic fields at $T_E\sim 130$-150 K, then increases drastically with the temperature lowering and saturates at liquid helium temperatures [50]. In the case of $CeCu_{6-x}Au_x$ [60] the transverse even effect is found to appear just below the structural phase transition at $T_s\sim 70$ K reaching maximum values in the quantum critical point $x_c$=0.1. The findings were discussed in [60, 50] in terms of anomalous Hall voltage caused by the interaction of external magnetic field with the filamentary structure of the many-body states. It is worth also noting that in the normal state of the HTSC in addition to the crystalline anisotropy the origin of the anomalous Hall signal is usually attributed to the appearance of stripes both on the surface and in the layers of the superconductors [18, 61]. A similar data analysis has been developed here for the configuration ($\mathbf{I}$//[1-10], $\mathbf{n}$//[110]) of the sample #1a, and it was found that both the amplitude of SIHE and the ordinary (cosine-type) terms are much smaller in comparison with the case ($\mathbf{I}$//[1-10], $\mathbf{n}$//[001]) (see Figs.10 and 4a). Moreover, the additional specific $\rho_H^{an}(\varphi)$ component in this configuration has several sign reversals with at least two intervals of negative values in the vicinity of the directions <111> (inset in Fig.10).

Fig. 11 demonstrates the decrease of both the additional Hall signal $\rho_H^{an}/H(T, H= 80$ kOe) and the field induced resistivity component $\Delta\rho(T, 80 \text{ kOe})= \rho(T, 80 \text{ kOe}) - \rho(T, 0 \text{ kOe})$ with the temperature increase (see Fig.3a). The behavior of these anomalous contributions may be well fitted by the analytical expressions $\rho_H^{an}/H= (\rho_H^{an}/H)_0 - AT^{-1}\exp(-T_0/T)$ and $\Delta\rho= \Delta\rho_0 - AT^{-1}\exp(-T_0/T)$ with the characteristic temperatures $T_0= 114-140$ K for $\rho_H^{an}/H$ and 140-150 K for $\Delta\rho$ (the fits are shown by the solid lines in Fig.11). Note, that the behavior may be discussed in terms of the formation of the fragments (chains) of the dynamic charge stripes below $T_E\sim150$ K, and then, the infinite cluster of the manybody states appears below $T^*\sim60$ K in the cage-glass state [62].

**3.6. HE and MR anisotropy for I//<100>.** In the previous part of the study we have focused only on the results of the charge transport measurements for the $\mathbf{I}$//[110] configuration, and the data obtained under the sample rotation around the $\mathbf{I}$//[110] current axis were discussed. Indeed, this configuration is the most important and useful for the analysis of the transverse ($\mathbf{I}\perp\mathbf{H}$) galvanomagnetic effects in LuB$_{12}$, because, from one side, the charge stripes are directed along <110> axes [30, 31], and, from the other, all three principal directions $\mathbf{H}$//[001], [1-10], and [1-11] are testified when rotating the crystal around this axis. At the same time, it is of extreme importance to investigate other $\mathbf{H}$-$\mathbf{I}$ configurations and to characterize in more details the charge transport anisotropy induced by the filamentary structure of quantum conduction channels in this heterogeneous medium. To check it both the transverse MR and Hall resistivity have been studied here with the current direction **I**//<100>. Figs. 12a and 12b show the angular (at T=4.2 K) and temperature dependences of resistivity, respectively, measured for the sample #2 in strong magnetic field (H=80 kOe). It is clearly discerned from Figs.12a-b, that below $T^*\sim60$ K the resistivity increases noticeably. Simultaneously, (*i*) two <110> directions in the (001) plane of the *fcc* LuB$_{12}$ lattice became non-equivalent and (*ii*) the $\rho(\varphi)$ dependence demonstrates a set of minima and maxima separated by $\Delta\varphi\approx22.5^o$. The polar plot presented in Figs.12c-d shows the variation of the angular MR dependences in magnetic field for the samples #2 and #1c of LuB$_{12}$, correspondingly. Usual 4-fold anisotropy of transverse MR is detected for sample #2 in the low magnetic fields $H\leq40$ kOe when the current is applied along the cube edge in the *fcc* lattice (Fig.12c). In higher magnetic fields the symmetry lowers dramatically when two <110> face diagonals become quite different. On the contrary, the 4-fold symmetry for the crystal #1c is found to break already for H$\geq20$ kOe (Fig.12d), so that two cubic edges [100] and [001] are happened to be non-equivalent. It is seen from Fig.13a that the dramatic symmetry breaking is developed in the sample #1c in the cage-glass state of LuB$_{12}$ (below $T^*\sim60$ K), so that 16 equidistant directions of magnetic field corresponding to the local maxima of transverse MR are detected in the sample.

It is also worth noting, that two [010]- and [001]-elongated bars of LuB$_{12}$ with equivalent (100), (010) and (001) faces were cut from one ingot in our previous study [29] (see, e.g., samples #1b and #1c in the inset in Fig.14a). At liquid helium temperatures their MR were found to be strongly anisotropic, demonstrating similar resistivity oscillations with the well-defined maxima separated by the period of $\Delta\varphi\approx22.5^o$ (Figs. 12a, 13a). We have also compared two sets of Hall effect data for cognate crystals #1b and #1c (Fig. 14) where the z axis is the same for both samples and magnetic field is applied transverse to the stripes (**H**//**n**//[001])**.** As a result, the reduced Hall resistivity in both crystals demonstrates similar anomalies in the wide neighborhood of the <001> direction. Note also that angular oscillations are also observed on the reduced Hall resistivity in addition to the main feature in the $\rho_H^{an}/H = f(\varphi, H=80$ kOe) curve, they being the most noticeable for the sample #1c (see the $\rho_H^{an}/H$ data in Fig.14b).

## 4. Discussion.

**4.1. Fermi surface (FS) effects.** At low temperatures the scattering of charge carriers in pure metals is sensitive to the topology of the Fermi surface (FS). Both quantum oscillations (de Haas-van Alphen (dHvA)/Shubnikov-de Haas (SdH) effects) and specific anomalies on the angular dependences of magnetoresistance and Hall resistivity are observed in the metals with open and closed electron orbits in high magnetic fields if $\omega_c \tau \gg 1$ ($\omega_c$ is the cyclotron frequency and $\tau$ is relaxation time) [56, 63-64]. In LuB$_{12}$ the dHvA signal has been measured at 0.35 K for all field directions within the (010) and (011) planes in the field range 9-12 T [38], but SdH oscillations at the same temperature have been found only for one combination of **I**//[110] and **H**//[001] [38]. In our measurements at T=4.2 K weak quantum oscillations have been also detected in magnetoresistance and Hall resistivity above 10 T under the same conditions (**I**//[110] and **H**//[001], see [54] for more detail). Note that both dHvA [65] and SdH oscillations depress totally when temperature increases over 10 K. So, even if the rough estimation of $\omega_c \tau$ ($\approx \mu_H H$) in the purest crystals of LuB$_{12}$ results in $\omega_c \tau < 1$ in magnetic fields below 12 T (see data for sample #1 in Fig. 4c) the anisotropy of the electron transport which could be induced by FS topology should be properly treated. Besides, the straightforward estimation of the concentration of conduction electron from high-temperature Hall constant $R_{H0} \approx -4 \cdot 10^{-4}$ cm$^3$/C (Fig.3) results in $n_0 \approx 1.56 \cdot 10^{22}$ cm$^{-3}$. The value in this metal with one conduction electron per formula unit [36-39] is considerably higher than the concentration of Lu ions ($n_{Lu} = 4/a^3 \approx 0.96 \cdot 10^{22}$ cm$^{-3}$, $a \approx 0.746$ nm is the lattice parameter [29-31,66]). This means that at least two groups of charge carriers (electrons and holes) with different effective concentrations and mobilities contribute effectively to charge transport in this non-magnetic metal at room and low temperatures.

This suggestion is supported by the FS structure of $LuB_{12}$ that consists from one hole and two electron sheets [38,39,65,67]. This "monster" in the $LuB_{12}$ lower conduction band [75] (see also [54]) is very similar to that one of the noble metals (Cu, Au and Ag) if allow for the hole-electron asymmetry of the related FS sheets [38]. Indeed, similar to copper the hole-like *open orbits* in $LuB_{12}$ are located close to <001>, <110> and <111> [56,63,64,68] while the nearest vicinity of all these axes provides only the *closed FS sections* for **H**//<001>, **H**//<110> and **H**//<111> (see [56,63,68] for more details). This feature results in the anisotropic MR behavior with twin peak structure around **H**//[001] and **H**//[110] and the saturation of $\rho/\rho_0 \approx 4$ for **H**//[111], which was detected in $LuB_{12}$ at T=0.5 K and $\mu_0 H$=12 T [38] (see also Fig. 5b-c) and was explained assuming several bunches of open orbits in the lower conduction band of this metal [38].

The similar topology of FSs in $LuB_{12}$ and Cu provides us by the explanation for the origin of the resistivity oscillations detected on the $\rho(\varphi)$ dependences for **I**//[010] (Figs. 12a, 13a). Indeed, the similar structure of the resistivity anomalies associated with the open orbits contribution was earlier detected for copper [63] (see polar plot Fig.12e). Note that the amplitude of these angular anomalies in the magnetoresistance of copper to be about 100-500 (Fig.12e) exceeds drastically this one in $LuB_{12}$ (~0.02-0.1, Figs. 12a-12d) due to higher quality of the Cu single crystals. Nevertheless, the similar positions of these resistivity maxima and the increase of their relative amplitude when increasing of magnetic field or/and lowering of temperature (Figs 12c-d, 13a) allow us to attribute these features to the contribution of open orbits, which appear for the particular directions of applied magnetic fields. In this approach, the peak features in magnetoresistance have to be accompanied by the minima in the Hall resistivity [56], which are transformed in the maxima under our subtraction procedure and can be clearly resolved in the angular dependences of the anomalous contribution $\rho_H^{an}/H$ (see arrows in Fig.14b).

At the same time, we have to note the principal dissimilarity in the MR behavior for copper and lutetium dodecaboride. In high quality crystals of copper the FS topology results in the MR saturation for all the principal directions of applied magnetic field (see deep minima in Fig. 5d), the unsaturated MR growth due to open orbits being detected close to all these axes. In LuB$_{12}$ the trend to $\rho$(H) saturation is observed for $\mathbf{H}$//[110] and $\mathbf{H}$//[111] only while MR does not saturate in the wide vicinity of $\mathbf{H}$//[001] (see Figs. 4b, 5a-5b and [38]). Note also, that the strongest MR increase in Cu is observed when $\mathbf{I}$//[110] and magnetic field is applied along the necks of the monster ($\mathbf{H}$//<111>) [54, 75]. Moreover, the MR anomaly at $\mathbf{H}$//[001] appears in LuB$_{12}$ at high temperatures (below $T_E\sim150$ K, see Figs. 11) even for the crystals with smaller mobilities (Fig. 5c and [54]). Finally, the position and the width of the MR anomaly in LuB$_{12}$ at $\mathbf{H}$//[001] are just identical to those ones of the extra specific contribution to Hall effect, which was extracted as $\rho_H^{an}/H$ in our study (see sections 3.5-3.6 and Fig.11).

In searching for the mechanism responsible for the appearance of the anomaly in MR and HE at $\mathbf{H}$//[001] note that the SdH quantum oscillations (see Fig.4 and [54]) were measured for $\mathbf{H}$//[001] at liquid helium temperatures in magnetic field $\mu_0 H > 10$ T in this metal with small enough average mobility of charge carriers (see $\omega_c\tau$ for sample #1 in Fig. 4c). This means that there is at least one group of carriers, for which the high-field limit $\omega_c\tau >> 1$ is valid. According to [38], the quantum SdH oscillations for $\mathbf{H}$//[001] are contributed from the hole-like orbits in the lower conduction band of LuB$_{12}$. This R$_{100}$ orbit ("four cornered rosette" [75]) corresponds to the particular dHvA $\alpha_2$ branch with the extreme area $S_{\alpha 2}\approx25$ nm$^{-2}$ [38]. If suggest that these hole-like regions are deformed spheroids in the W points of the Brillouine zone [38] one can estimate the effective parameters of these charge carriers. In particular, leaving for the geometrical factor $G=V_{\alpha 2}/S_{\alpha 2}^{3/2}=0.9$ ($V_{\alpha 2}$ and $S_{\alpha 2}$ – the volume and the extreme area of the FS sheet) to be intermediate between $G=1$ for cuboid and $G=4/(3\pi^{1/2})\approx0.75$ for spheroid, the volume of the FS sheet can be estimated as $V_{\alpha 2}=0.9S_{\alpha 2}^{3/2}\approx113$ nm$^{-3}$, which corresponds to the $n_{\alpha 2}\approx2.7\cdot10^{21}$ cm$^{-3}$ and the effective Hall constant $R_H^{\alpha 2}=(|e|n_{\alpha 2})^{-1}\approx+2.3\cdot10^{-3}$ cm$^3$/C. The condition $R_H^{\alpha 2}(\sigma_{\alpha 2})^2 \sim R_{H0}(\sigma_0)^2$ results in the conductivity ratio $\sigma_{\alpha 2}/\sigma_0\approx0.42$ ($\sigma_{\alpha 2}$ and $\sigma_0$ are the hole conductivity and the average conductivity, respectively). The corresponding mobility ratio is estimated as

$\mu_{\alpha2}/\mu^*=(\sigma_{\alpha2}n_0)/(\sigma_0 n_{\alpha2})\approx2.4$ resulting in the value of $\mu_{\alpha2}\approx7100$ cm$^2$V$^{-1}$s$^{-1}$. Thus, for these hole-like orbits the high-field limit $\omega_c\tau\sim\mu_{\alpha2}H_0\sim2\pi$ gives the reasonable value for $\mu_0H_0\sim9$ T that agrees very well with the onset of SdH oscillations on the $\rho(H)$ and $\rho_H(H)$ data for $\mathbf{H}//[001]$ (Fig.4 and [54]). The estimation of relaxation time results in the value of $\tau=\mu_{\alpha2}m^*_{\alpha2}/|e|\approx2$ ps if allow for the effective mass value $m^*_{\alpha2}\approx0.53m_0$ [38]. The corresponding Dingle temperature $T_D\approx0.6$ K and the high enough value of related parameter $\mu_0H\tau\approx16$ T·ps [65] prove these quantum oscillations to be observable at liquid helium temperatures.

The above speculations allow to suggest that the wide anomaly detected at $\mathbf{H}//[001]$ in the angular dependences of MR and Hall resistivity may be induced by a strong relaxation time anisotropy rather than by any specific contribution of the open orbits on the LuB$_{12}$ FS, which seem to produce only narrow anomalies on the MR data in available magnetic fields (see Fig.5b-5c and [38]). The anisotropy of scattering time is more pronouncing in Hall effect data where noticeable deviations from cosine behavior are well resolved below 150 K (Fig.9c,e). It should be noted here that these temperatures are comparable to the vibrational energies of the rare earth ions (scaled by Einstein temperature $T_E\approx150$K) so as strong electron-phonon interaction has to smear out any FS induced features. In this respect, any consistent model allowing for scattering time anisotropy should comprise the microscopic mechanisms, which are responsible for the formation of these "hot" and "cold" regions of FS, and the related factors, which prevent any effective mixing of these different electron states.

**4.2. Stripe induced Hall effect (SIHE).** The above consideration shows that any model based exclusively on the particular FS topology is insufficient to explain not only the origin of the MR anomaly for $\mathbf{H}//[100]$ and $\mathbf{I}//[011]$ (Figs. 5b-c, 6b), but also the extremely narrow step-like feature $\Delta_{step}$ and a wide range $\Delta_{smooth}$ component detected in the angular dependences of Hall resistivity at $\mathbf{H}//[001]$ for all directions of applied current (Figs.7, 9, 14b). Indeed, the wide MR anomaly at $\mathbf{H}//[001]$ is observed even for the crystals of LuB$_{12}$ with a small enough mobility (Fig. 5c and [54]) and at high temperatures up to 150 K (Fig. 11). It should be point out that the step-like feature in Hall resistivity at $\mathbf{H}$//[001] is detected only for one particular combination of $\mathbf{H}$//[100] and $\mathbf{I}$//[011] in the highest quality crystals of LuB$_{12}$ at temperatures up to $T^*\sim$60K (see Fig.9c). In this respect we note that the effects of symmetry lowering were previously established in the MR study of LuB$_{12}$ (Fig.12-13) as the unique features of the RE dodecaboride with the dynamic charge stripes. Indeed, the corresponding electron density (ED) distribution in the LuB$_{12}$ crystals, which was deduced from the analysis of the x-ray diffraction data with the help of the maximal entropy method (MEM) [30], is strongly inhomogeneous (Fig.13b). Assuming that pinning of the dynamic charge stripes on the randomly arranged defects and impurities fixes the selected direction in the dodecaboride matrix (Fig.13b), it is naturally to associate the *fcc* symmetry breaking with the nano-scale phase separation effects and emerging filamentary structure of conduction channels in the LuB$_{12}$ matrix.

Note here that the discussed anomalies found in the angular dependences of magnetoresistance and Hall resistivity are specific for both magnetic and nonmagnetic Ho$_x$Lu$_{1-x}$B$_{12}$ compounds [70, 71]. These common features increase strongly below $T^*\sim$60 K (see Fig.11) in the cage-glass state [40, 29]. In our opinion, these similarities favor to the explanation for the significant MR increase in the vicinity of $\mathbf{H}$//[001], which has been proposed recently in [70, 72]. We consider the strong interaction of the dynamic stripes along the [110] direction with the transverse external magnetic field (Fig.1a) as the possible reason for the detected increase of the amplitude of the negative cosine-type component (absolute values of the ordinary Hall effect), which is clearly observed with increase of external magnetic field (Fig. 4a). This behavior (see Figs. 7-8 and Figs.S4,S5 in [54]) is accompanied by the changes of the average mobility in the LuB$_{12}$ crystals. The measurements of optical conductivity in LuB$_{12}$ [35] revealed that the conduction band consists of (*i*) Drude electrons and (*ii*) non-equilibrium (hot) charge carriers involved in the collective modes and seen in the conductivity spectra as overdamped oscillators. As non-equilibrium electrons amount about ~70% of the total electron concentration, the related approach should include some redistribution of electrons between these two electron sub-systems. A rough estimation obtained from the optical conductivity spectra [67] proves that the relative number of the Drude-type and hot charge carriers does not changed in $LuB_{12}$ at low temperatures. We suggest that strong impurity scattering provokes fast thermalization of the hot electrons, but, on the contrary, strong external magnetic field increases the number of electrons in these dynamic charge stripes. As a result, both the impurity concentration and external magnetic field may be considered as two factors, which are responsible for the redistribution of charge carriers between the non-equilibrium and Drude-type electron sub-systems in the heterogeneous medium of $LuB_{12}$ non-magnetic metal with dynamic charge stripes.

This redistribution may also increase the amplitude of the Hall effect component, which is induced by the stripes (SIHE). In this scenario the low temperature enhancement of the negative ordinary Hall coefficient (marked as *cos* in Fig. 4a) observed in strong magnetic field should be attributed to the decrease of free electron concentration $n_e$. The $n_e$ evolution is accompanied with a dramatic increase of the SIHE amplitude (Figs. 9d, 9f, 11 and 14). In this picture the field-induced decrease of a number of carriers $\Delta n_e/n_e$ in the $LuB_{12}$ matrix is dependent on the direction of external magnetic field. The effect is happened to reach the largest value (~30%) for **H** to be transverse to dynamic charge stripes (**H**//**n**//[001], see schema in Fig.1a), while only relatively small (~10%) changes are observed for **H**//[110] (Fig. 4a). The validity of the SIHE scenario could be verified in future both by precise magnetotransport study and optical measurements in high magnetic field. Note here only that the redistribution in magnetic field of the charge carriers between the Drude-type and hot, non-equilibrium electron sub-systems may be attributed to the reasons of emergence of anomalous Hall effect in the dodecaborides with magnetic RE ions, e.g., such as $Ho_xLu_{1-x}B_{12}$ [76].

## 5. Conclusion.

Summarizing the specific features of magnetoresistance and Hall resistivity found in the present study for the non-magnetic metal $LuB_{12}$ with dynamic electron stripes, we outline the different factors, which may result in the anomalies of electron transport in this material. While some peculiarities found in the angular dependences of MR (16 maxima for I//[001] and twin peak fine structure around **H**//[001] and **H**//[1-10] for I//[110]) and Hall resistivity may be definitely associated with the presence of open orbits in the corresponding sections of FS, the origin of the wide feature at H//[001] found both in MR and HE requires further theoretical consideration. The estimations based on the FS topology show that this anomaly could appear due to the relevant contribution of the holes with large enough mobility (7100 $cm^2V^{-1}s^{-1}$) and relaxation time (2 ps), but these parameters decrease dramatically at high temperatures (up to 150 K) prohibiting this kind scenario. In contrary, the inhomogeneous distribution of electron density in $LuB_{12}$ initiated by the cooperative JT effect in the $B_{12}$ clusters in the presence of the rattling modes of Lu ions may also result in the additional component in Hall resistivity. In the last case the configuration dependent contribution to Hall effect arises due to interaction of magnetic field with the filamentary structure of dynamic charge stripes, which pin occasionally along one of <110> principal directions of the *fcc* lattice. However, the approach requires a careful consideration of the processes, which determine the redistribution of the charge carriers between "cold" (Drude-like) and "hot" (non-equilibrium) channels. Indeed, the strong anisotropy of relaxation time with its low enough average value may be induced by the cooperative JT instability of the $B_{12}$- clusters in $LuB_{12}$. The related ferrodistortive effect in the boron sublattice may induce the periodic changes in the $5d$-$2p$ hybridization modulating the width of the conduction band and increasing the intensity of charge carrier scattering. In this scenario the related "breathing" of FS should lead to periodic changes of size for both the belly and necks of the hole-like FS monster [54] and may be considered as the main factor responsible for the average $\tau$ decrease and the depression of the SdH oscillations in $LuB_{12}$. In this respect the step-like feature in Hall resistivity detected only for one combination of **H**//[100] and **I**//[011] in the high quality single crystals of $LuB_{12}$ (see sections 3.5-3.6) may be treated as a key feature to check the validity of the theoretical approaches proposed for explanation of the magnetotransport anomalies in this compound.

**Acknowledgements.** This work was supported by the Russian Science Foundation, project no. 17-12-01426 (MR measurements) and Russian Foundation for Basic Research, project no. 18-02-01152 (Hall effect study). The authors are grateful to V.N. Krasnorussky for experimental assistance and useful discussions.

**References.**

resistivity for the sample #2 of LuB$_{12}$ (**Figs.S7-S8**); the hole-type Fermi surface sheet in LuB$_{12}$ (**Fig.S9**); quantum oscillations of resistivity and the reduced Hall resistivity in LuB$_{12}$ (sample #1a) (**Fig.S10**).

**Figure captions.**

**Fig.1.** (a) Crystal structure of $RB_{12}$. Color plane shows the map of the electron density in the face (001) obtained by the maximum entropy method (MEM) [29]. Green bars correspond to the dynamic charge stripes. Directions along (**n**//[110]) and transverse (**n**//[001]) to the stripes are shown. (b) Fragment of the crystal structure demonstrates two connected $B_{24}$ clusters with embedded $Lu^{3+}$ ions. (c) The schematic view of two double-well potentials for neighboring Lu-ions.

**Fig. 2.** Temperature dependences of (a) the resistivity and (b) Hall mobility at H=80 kOe of the $LuB_{12}$ crystals. Vertical dashed line marks the transition at $T^*{\sim}60$ K to the cage-glass state.

**Fig.3.** Temperature dependences of resistivity for the $LuB_{12}$ crystals #1a (panel a), #2 (c) and #3 (e), and of the reduced Hall resistivity for crystals #1a (panel b), #2 (d) and #3 (f) at H=80 kOe. Vertical dashed line marks the transition at $T^*{\sim}60$ K to the cage-glass state. Large open symbols correspond to the ordinary Hall coefficient (marked as cos) and the sum of the ordinary and anomalous components (see text for details).

**Fig.4.** Magnetic field dependences of the reduced Hall resistivity (a), resistivity (b), and the Hall mobility (b) of the $LuB_{12}$ crystal #1a at the liquid helium temperature. Large open symbols on the panel (a) correspond to the ordinary Hall coefficient (marked as cos) and the sum of the ordinary and anomalous components (see text for details).

**Fig.5.** Magnetic field (a) and angular (b) dependences of resistivity of the sample #1a of $LuB_{12}$ at T=4.2 K. For comparison the resistivity $\rho(\varphi, T=4.2$ K, $H=80$ kOe) curves for several $LuB_{12}$ samples are shown in panel (c). Schematic view of the sample for the galvanomagnetic measurements is presented in the insert in panel (a). Panel (d) shows the results of similar transverse MR rotations for copper ($T=4.2$ K, $H=18$ kOe, RRR=8000, adapted from [63]).

**Fig.6.** (a) The polar plot of resistivity of the sample #1a at T=4.2 K. (b) The presentation in the cylindrical coordinates of the temperature dependence of MR of the sample #1a at H=80 kOe. Color shows the MR amplitude.

**Fig.7.** Angular dependences of the reduced Hall resistivity $\rho_H/H$ for the samples #1a (panel a), #2 (b) and #3 (c) at liquid helium temperature in the configuration **I**//[110], **n**//[001]. Solid line in

(c) demonstrates the cosin-type approximation. Vertical arrows in (a) show two H directions in the traditional, common-used technique of the Hall effect measurements.

**Fig.8.** Angular dependences of the reduced Hall resistivity $\rho_H/H$ for the samples #1a (panel a), #2 (b) and #3 (c) at liquid helium temperature in the configuration **I**//[110], **n**//[110]. Vertical arrows in (a) show two **H** directions in the traditional, common-used technique of the Hall effect measurements.

**Fig.9.** Angular dependences of the reduced Hall resistivities $\rho_H/H$ and $\rho_H^{an}/H$ in the configuration **I**//[110], **n**//[001] in the sample #1a for various temperatures at $H$=80 kOe (panels a and c) and in different magnetic fields at T=4.2 K (panels b and d), correspondingly. Solid lines in (b) demonstrate the examples of the cosin-type approximation. $\Delta_{smooth}$ and $\Delta_{step}$ in (c) and (d) denote two anomalous components in the anomalous reduced Hall resistivity $\rho_H^{an}/H$. The reciprocal temperature plot in (e) shows the behavior of $\rho_H^{an}/H$=$f(T)$ detected in the crystals #1a and #2. Panel (f) presents the field dependence of $\rho_H^{an}/H$=$f(H, T$=4.2 K) and $\Delta_{smooth}$, $\Delta_{step}$ components. $T_E$ and $T^*$ are the energy scales in the LuB$_{12}$ (see text).

**Fig.10.** Angular dependences of the reduced Hall resistivities $\rho_H/H$ and $\rho_H^{an}/H$ in the configuration **I**//[1-10], **n**//[110] in the sample #1a for various magnetic fields at $T$=4.2 K. Solid lines demonstrate the examples of the cosin-type approximation. Insert shows large scale view of $\rho_H^{an}/H$=$f(\varphi, H$=80 kOe, $T$=4.2 K).

**Fig.11.** The temperature dependences of the anomalous reduced Hall resistivity $\rho_H^{an}/H$ and resistivity $\Delta\rho$ components for the samples #1a (a) and #2 (b). Solid lines show the analytical approximation by $(\rho_H^{an}/H, \Delta\rho)$= $(\rho_H^{an}/H, \Delta\rho)_0$ - $AT^{-1}$exp(-$T_0/T$). Vertical dashed line marks the transition at $T^*$~60 K to the cage-glass state.

**Fig.12.** (a) Angular (T=4.2 K) (b) temperature dependences of resistivity of the sample #2 of LuB$_{12}$ at $H$=80 kOe. For comparison, the polar plots (c) and (d) at $T$=4.2 K demonstrate the angular dependences of the normalized MR for the samples #2 and #1c, correspondingly). Panel (e) shows the results of similar resistivity/transverse MR rotations for copper ($T$=4.2 K, $H$=18 kOe, RRR=8000, adapted from [77]).

**Fig.13.** (a) The presentation in the polar coordinates of the temperature dependence of resistivity of the sample #1c at H=80 kOe (**I**//[010]). Color shows the resistivity amplitude, arrows mark the harmonics (see text). (b) The map of the electron density distribution in $LuB_{12}$ obtained by the maximum entropy method [29].

**Fig.14.** Angular dependences of the reduced Hall resistivity $\rho_H/H$ for the samples #1b (panel a, $\mathbf{I}//[100]$) and #1c (b, $\mathbf{I}//[010]$) at liquid helium temperature in the configuration $\mathbf{n}//[001]$. Solid line in (b) demonstrates the cosine-type approximation, $\rho_H^{an}/H(\varphi)$ shows the anomalous component. Arrows marks the harmonics (see text). Insert shows schematically how the samples #1b and #1c cut from the single crystalline disc of $LuB_{12}$.

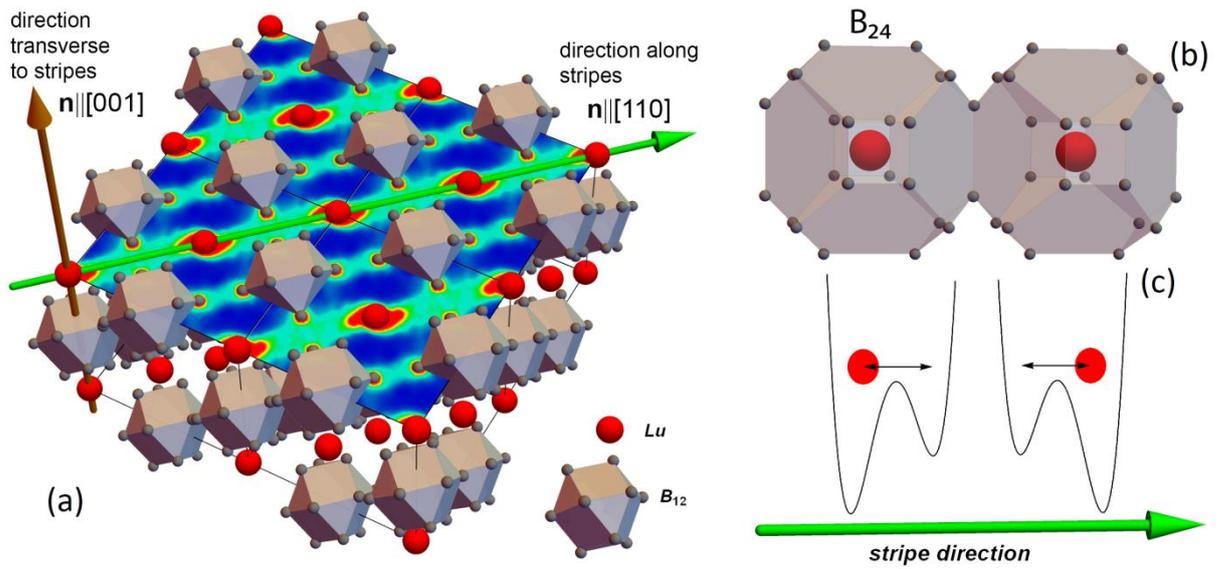

**direction transverse to stripes** n‖[001]

**direction along stripes** n‖[110]

$B_{24}$ (b)

(c)

*stripe direction*

Lu

$B_{12}$

(a)

**Fig.1.**

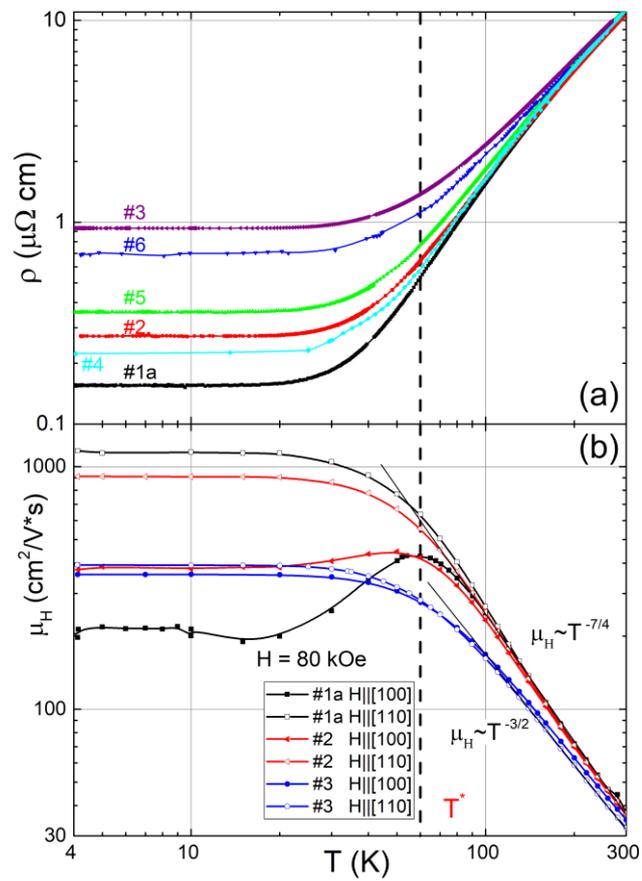

**Fig. 2.**

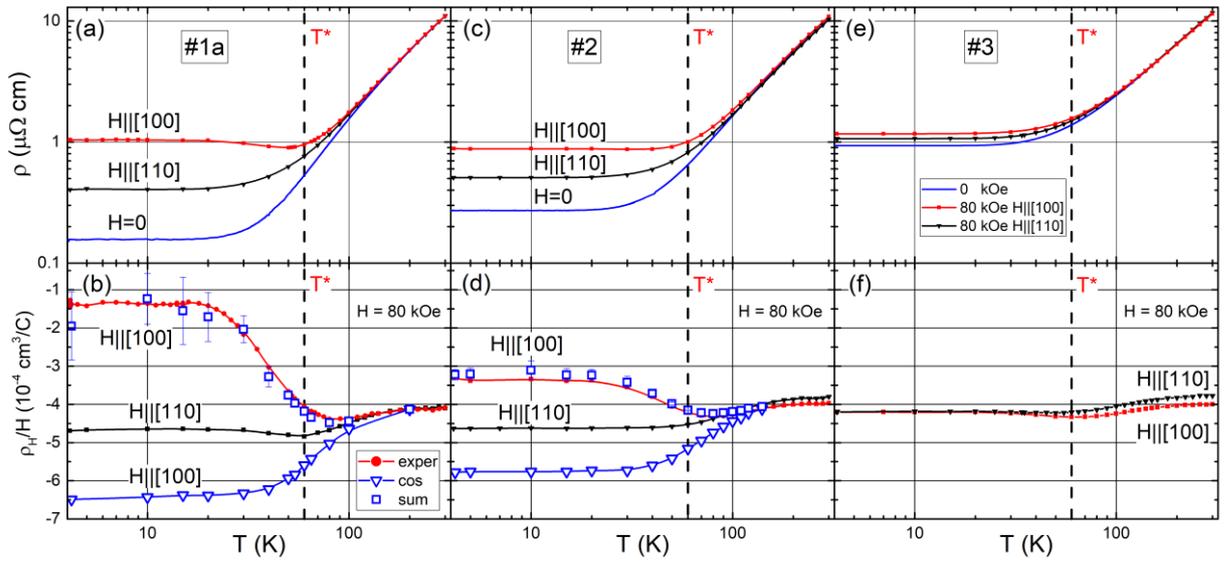

Fig.3.

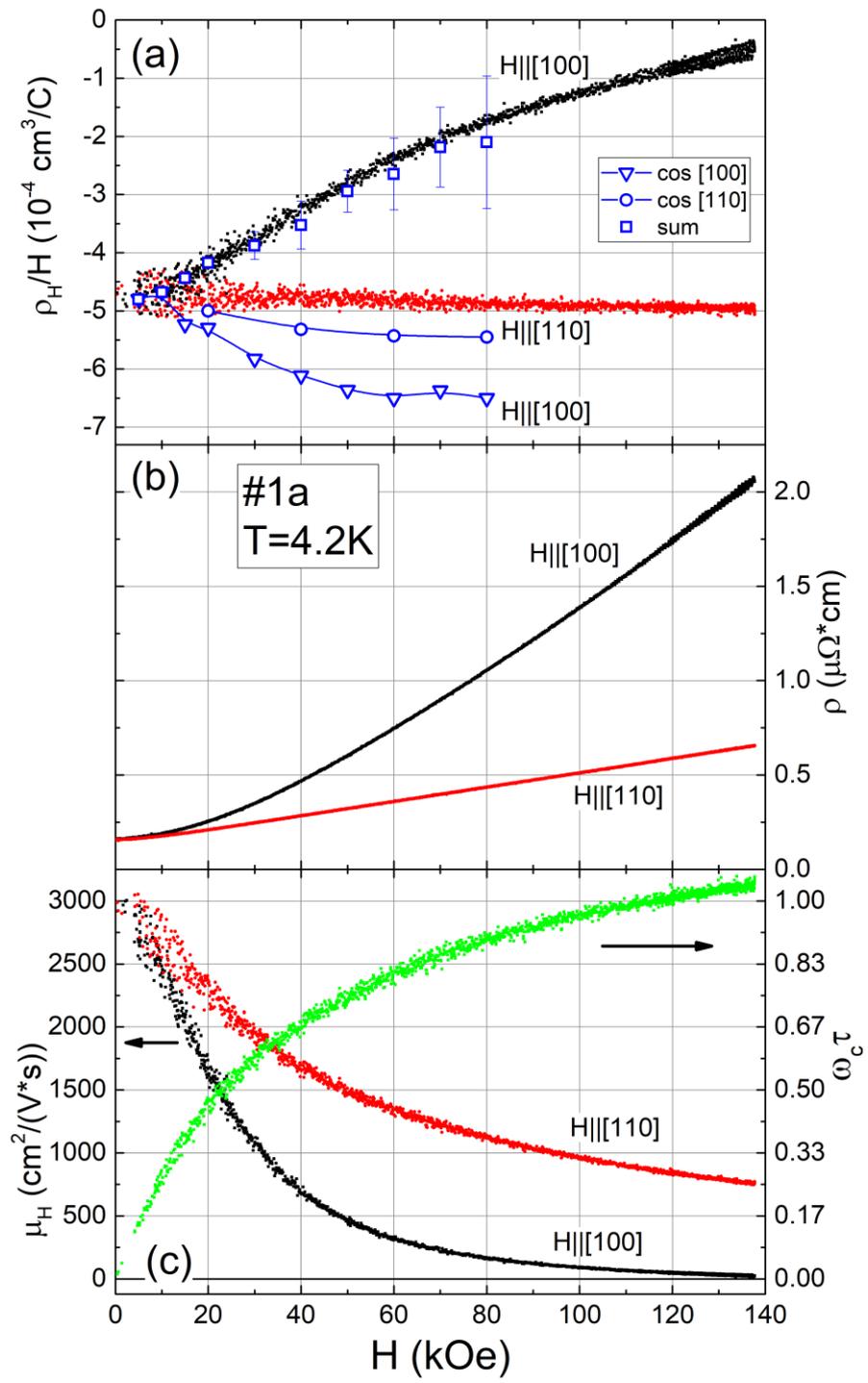

Fig.4.

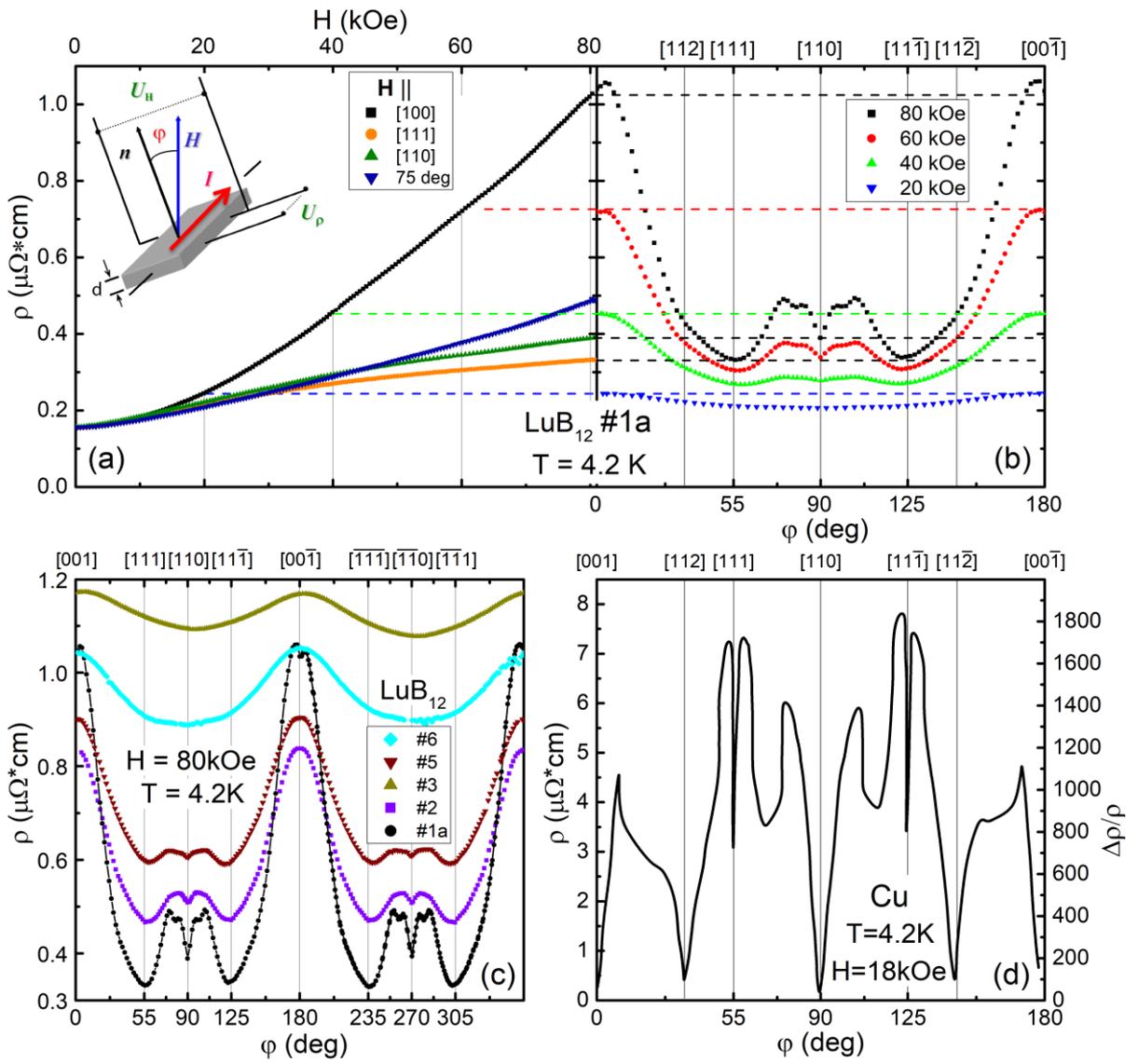

**Fig.5.**

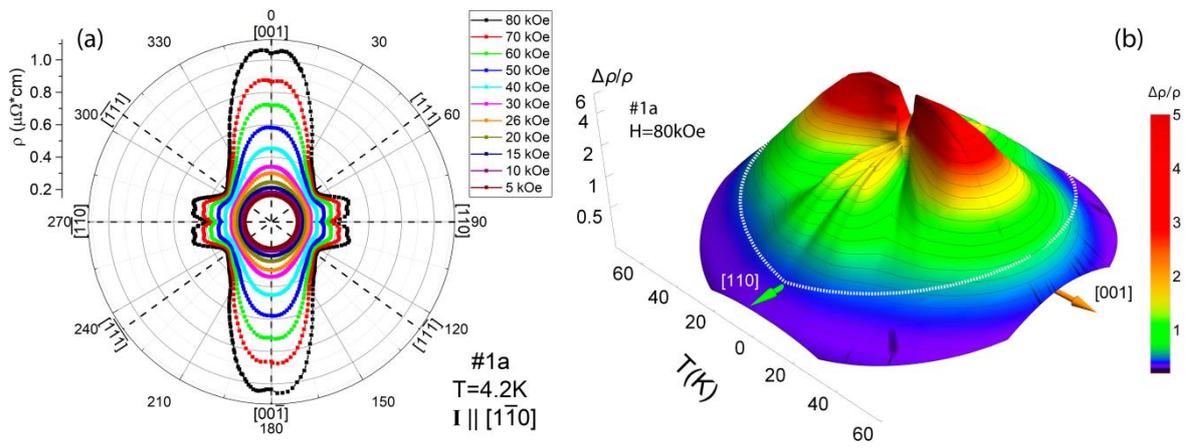

**Fig.6.**

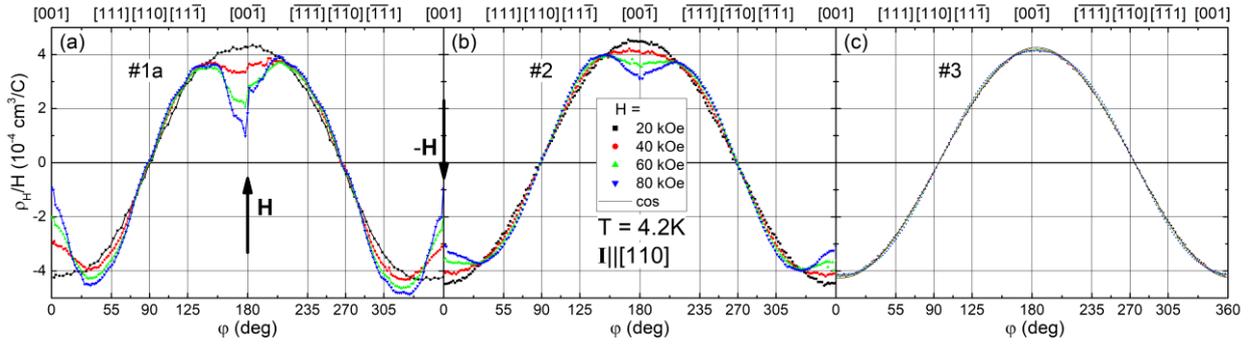

**Fig.7**

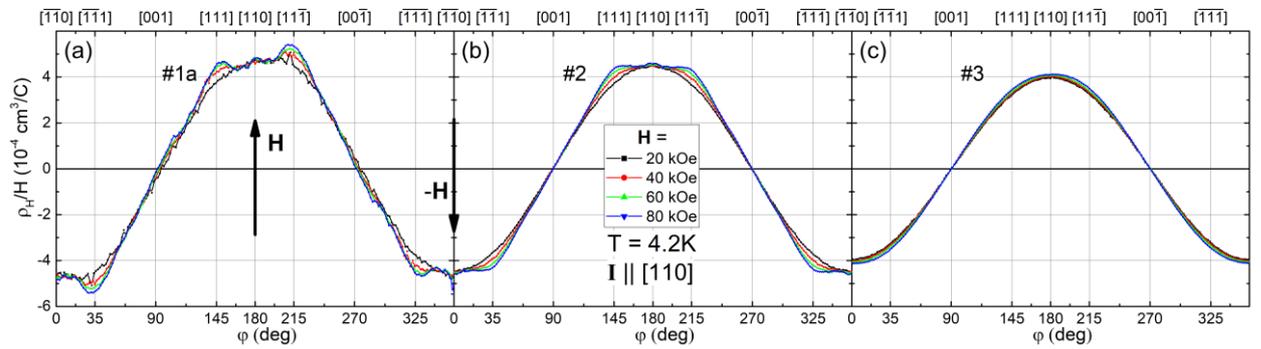

**Fig.8.**

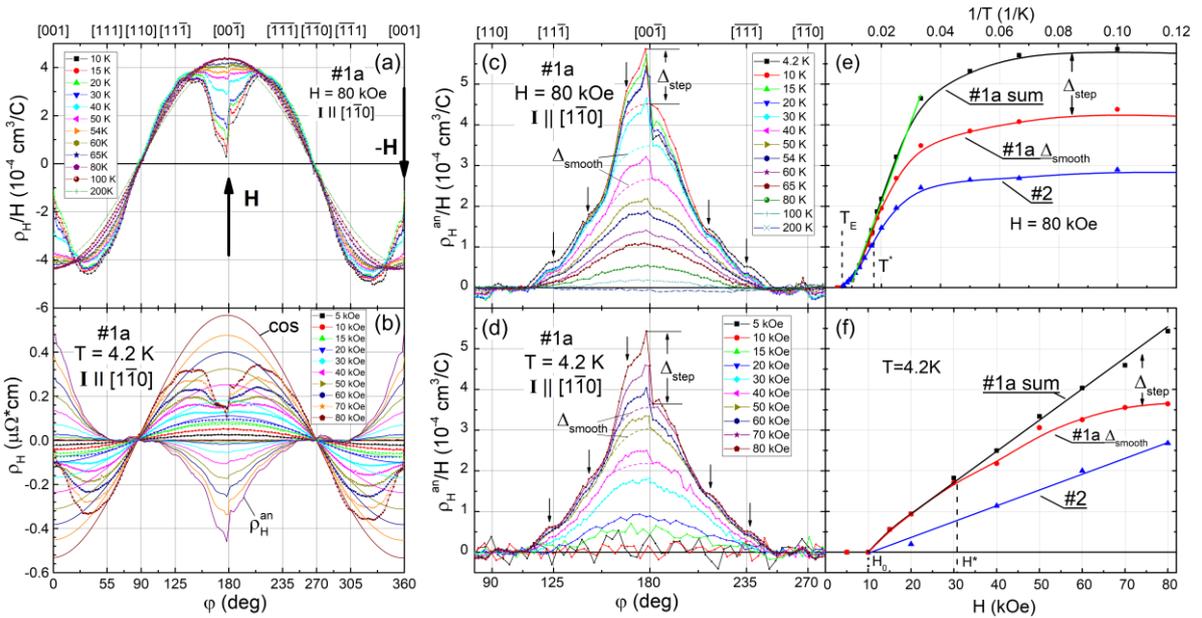

**Fig.9.**

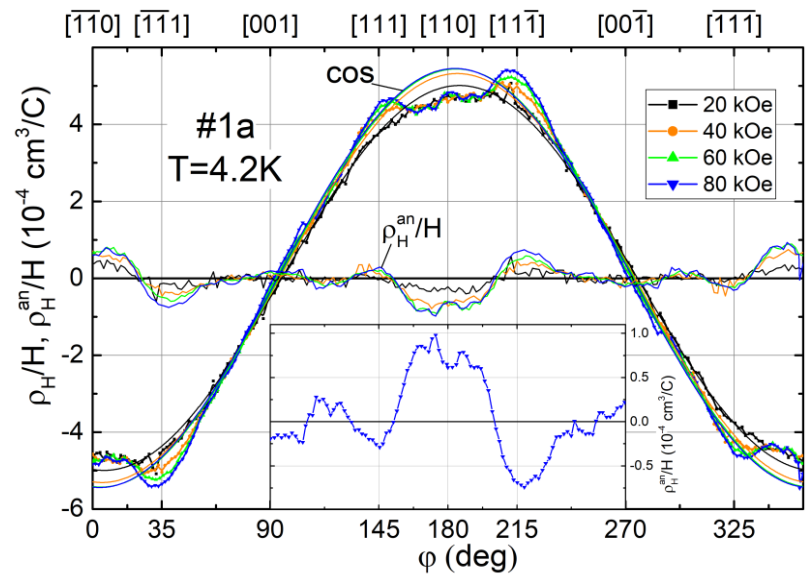

**Fig.10.**

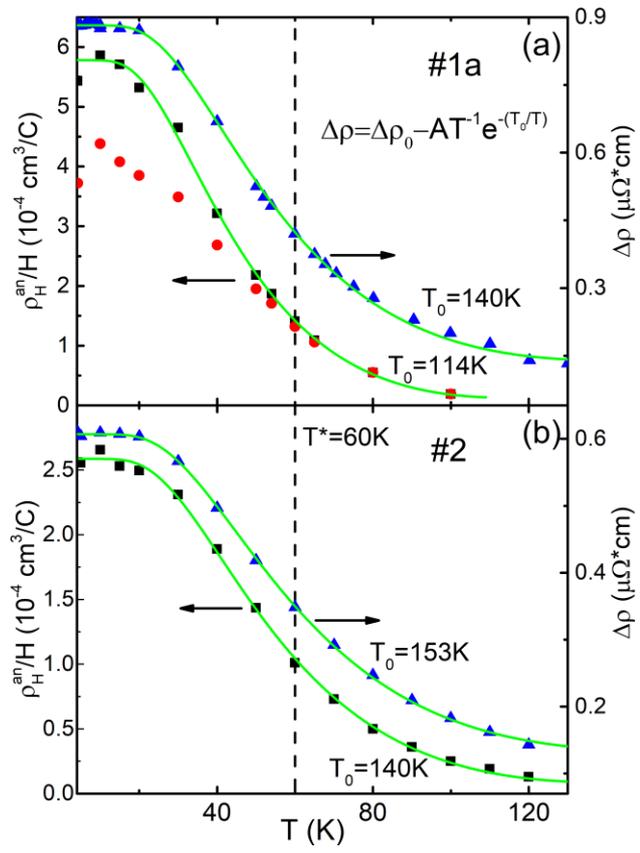

**Fig.11.**

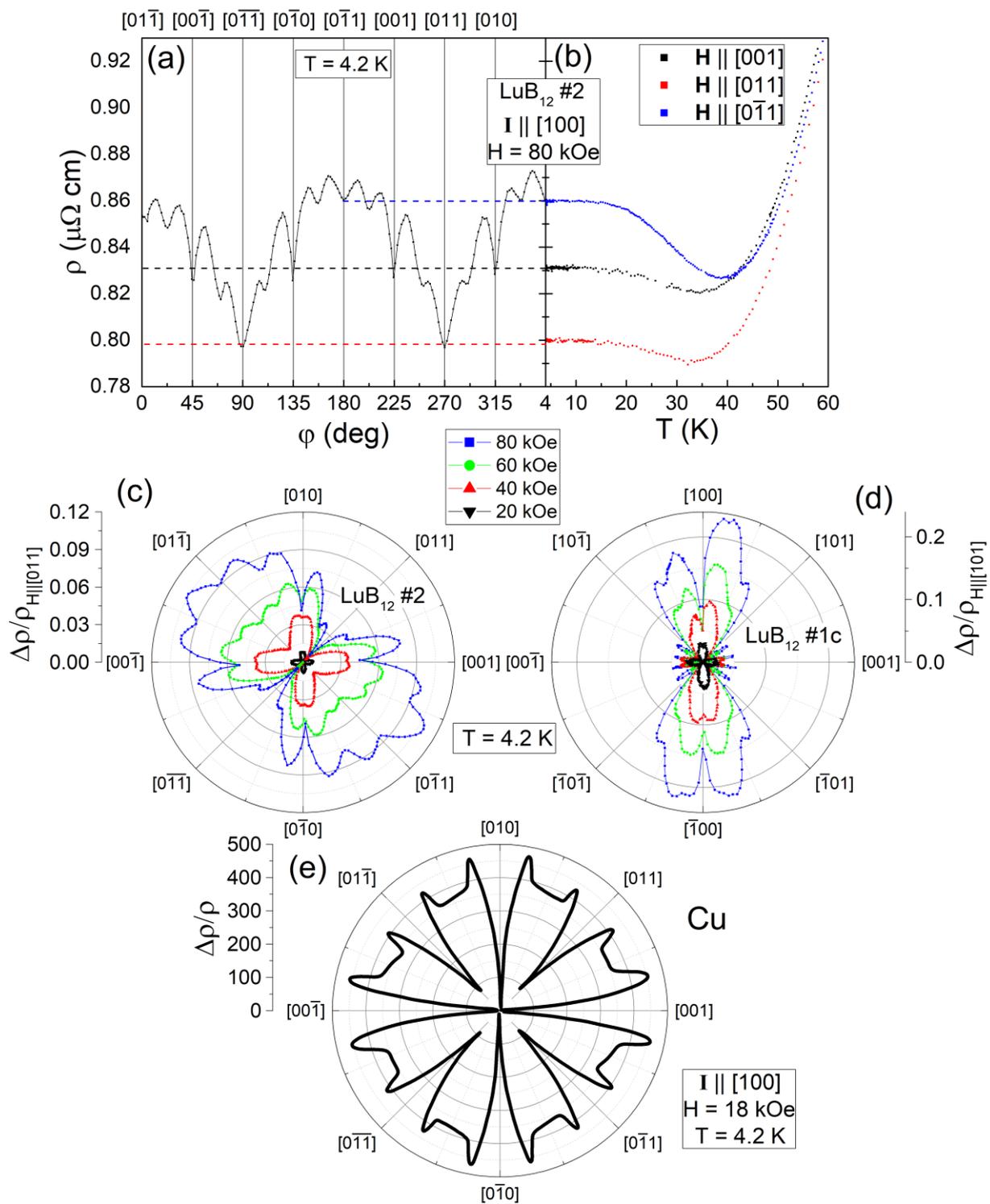

Fig.12.

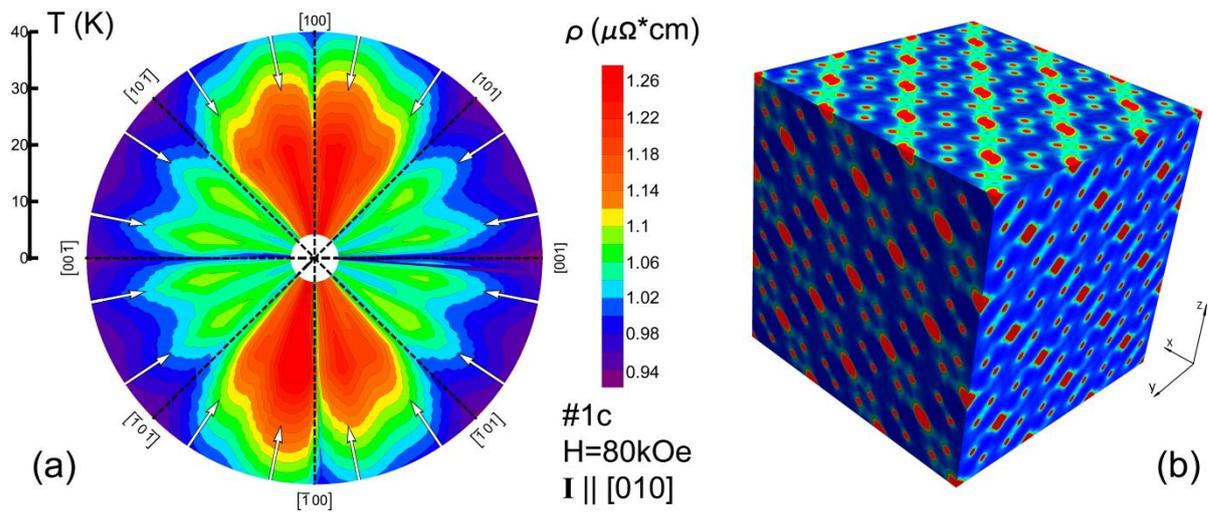

**Fig.13.**

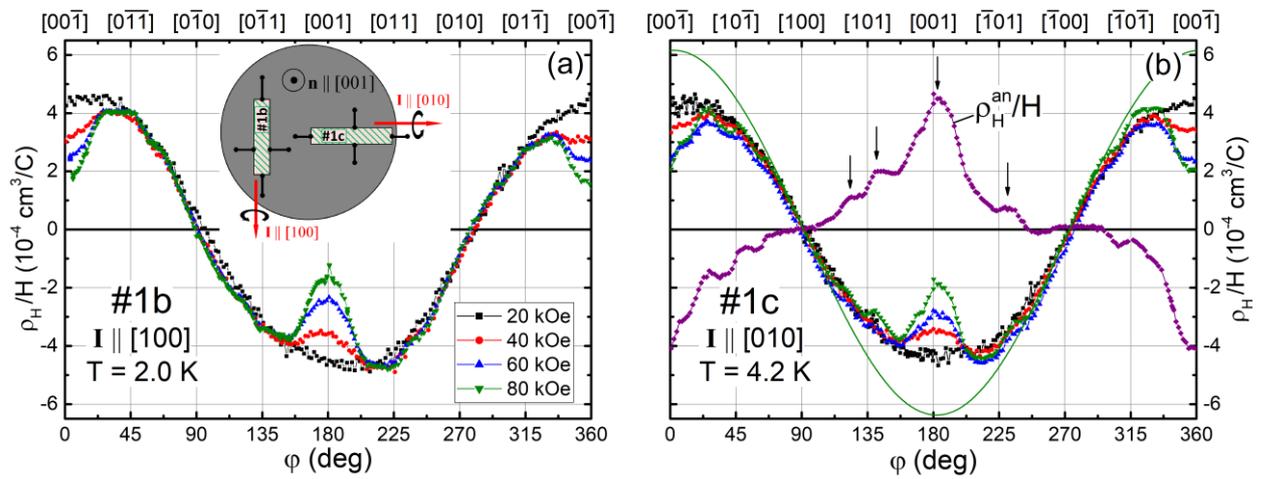

**Fig. 14.**

# Supplementary Information

to the paper of N. Sluchanko, A. Azarevich, A. Bogach, S. Demishev, K. Krasikov, V. Voronov,

V. Filipov, N. Shitsevalova, V. Glushkov

**"Hall effect and symmetry breaking in non-magnetic metal with dynamic charge stripes".**

### 1. The crystal growth.

$Lu_2O_3$ powder was preliminarily annealed at $800^0$C for 2 hours to remove the crystallization water, after that the charge was prepared from oxide and boron according to the equation of a solid-state reaction of the borothermal reduction:

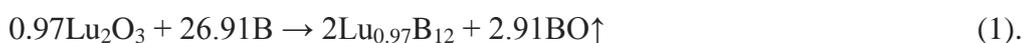

$$0.97Lu_2O_3 + 26.91B \rightarrow 2Lu_{0.97}B_{12} + 2.91BO\uparrow \qquad (1).$$

The prepared charge was mechanically mixed for several days, no less than 5 times being sifted through a sieve in order to break conglomerates based on oxide and boron. It was necessary to prepare, as far as possible, a homogeneous mixture. The prepared mixture was compressed into tablets with a diameter of 15 mm and a height of 10 mm, which were aged in a vacuum furnace for an hour at a temperature of $1650^0$C. Since the solid-phase synthesis is determined by the diffusion of boron into oxide particles, then to obtain a homogeneous composition, the sintered tablets were broken, pressed again and kept in vacuum at $1750^0$C for an hour. The annealed tablets were again broken; the powder was cold pressed into rods with a diameter of 8 mm and a length of 60 mm and sintered for one hour at $1750^0$C in vacuum. The final homogenization occurs in the melt in the process of crystal growth.

The synthesis, annealing of tablets and sintering of rods took place in the same crucibles from $ZrB_2$. The melting temperature of $ZrB_2$ is ~ $3000^0$C; therefore, contamination with crucible material is excluded. Thus, the initial $LuB_{12}$ rods for crystal growth were identical in composition and purity. Volatile impurities present in boron are removed during the synthesis and zone melting, and in the grown crystals according to the optical spectral analysis the amount of impurities does not exceed $10^{-3}$ mass % (except for RE); rare earth impurities are determined by the purity of the starting lutetium oxide, the total content of the concomitant RE impurities is not more than $1.5 \ 10^{-3}$ %.

Zone melting is carried out in a closed chamber under the pressure of high-purity argon (volume fraction of argon is not less than 99.993 %). The process of the chamber preparing for melting is also identical for all crystals. Preliminarily the chamber is pumped out to $10^{-3}$ mm Hg (0.1333 Pa), then it is filled with argon up to 0.2 MPa and again is pumped out, after which it is filled with argon to a predetermined pressure, and the melting process begins.

In the case of the crystal No. 1, an additional purification of argon from possible impurities of other gases was carried out directly in the growth chamber, whose presence cannot be excluded given their partial pressures, whatever the quality of the system sealing. Oxygen is especially undesirable as it promotes the growth of block crystals. Ti was used as a getter. In addition, during the growth of this crystal, both the feeding rod and the growing crystal rotated in opposite directions at rates of 10 and 4 rpm, respectively, which favored to a more uniform temperature distribution in the melt zone and equalization of the crystallization front, reducing thermal and mechanical stresses in the crystal. This crystal was grown with one pass, under a pressure of Ar 0.5 MPa, the seed was fixed in the upper shaft, and linear velocities were 0.5 and 0.4 mm/min for the growing crystal and the feeding rod respectively. The Laue photo (not presented) demonstrated the absence of split-in of the reflexes, i.e. a mosaity structure was not detected and a quality single crystal was grown.

Over time, we refused from additional gas purification, since the split-in of reflexes at the Laue photos of the grown crystals was absent as before.

The remaining crystals, with the exception of crystal #3 with three zone passes, were also grown with one pass. The crystallization rates of all crystals differ slightly.

Crystals #2 and #4 were grown under identical conditions, the seed was fixed in the lower shaft, only the feeding rod rotated (5 rpm), the crystallization rate was slightly lower (0.45 mm/min), but the pressure in the chamber was reduced (0.2 MPa), in order to facilitate purification from volatile impurities from boron, and a piece of sintered boron was added to the initial zone to somewhat lower the temperature of the melt zone and achieve a more stable growth process. According to [S1] the boron predominant evaporation resulted in a shift of the composition within the homogeneity region to a side rich in lutetium, i.e. the formation of defects in the boron sub-lattice, which led to an increase in residual resistance. This did not affect the quality of the Laue photos.

In the case of sample #5, the pressure in the chamber was further reduced (P = 0.15 MPa), and, accordingly, the concentration of defects in the boron sub-lattice was increased.

Crystal #3 was melted three times, the pressure in the chamber was 0.2 MPa at each pass, the rate of movement of the crystallizing ingot was 0.45 mm/min, giving 0.4, 0.42, 0.42 mm / min, and its rotation rate was 3, 5, 5 rpm / min., respectively. The grown ingot was much stressed, the end of the zone was in small cracks, but the Laue photo at a distance of 5 mm from the end of the crystal zone was very good. Thus, mechanical stresses, as a result of deviations from stoichiometry caused by three-fold re-melting and temperature gradients at the melt-to-crystal interface worsened the real structure of the crystal and increased residual resistance, but this did not affect the macrostructure.

Thus, indirectly, based on the values of the residual resistance, the influence of the production conditions on the quality of the grown crystals is detected. Unfortunately, the error in the determination of boron concentration by analytical methods is very large and it was not possible to determine the exact composition of the grown crystals.

## 2. Supplementary information to the results of the charge transport in LuB₁₂.

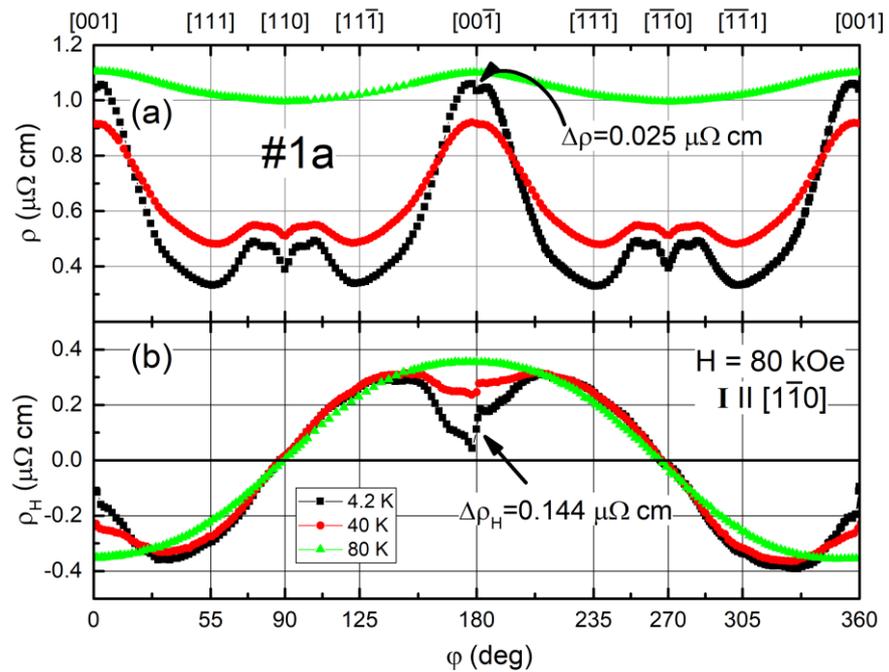

**Fig.S1.** Comparison of the amplitude of the resistivity and the Hall resistivity anomalies on the angular dependences near **H**//**n**//<001> in LuB₁₂ (#1a) at the temperatures 4.2-80 K.

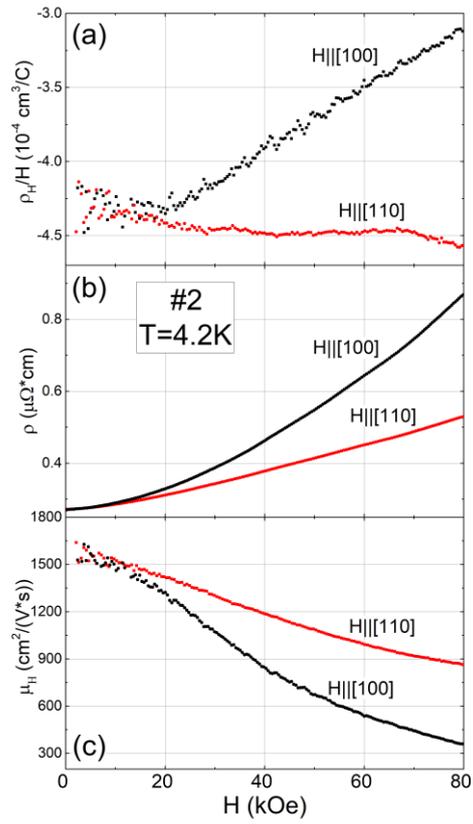

**Fig.S2.** Magnetic field dependences of (a) the reduced Hall resistivity $\rho_H/H$, (b) resistivity and (c) Hall mobility in LuB$_{12}$ (#2) at T=4.2 K.

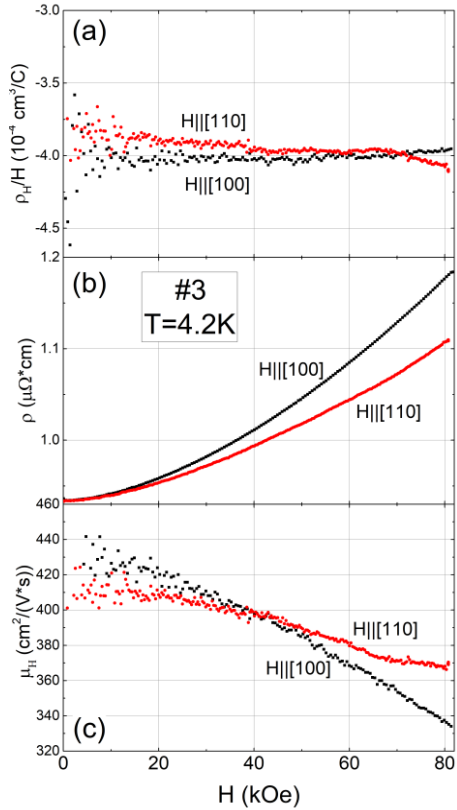

**Fig.S3.** Magnetic field dependences of (a) the reduced Hall resistivity $\rho_H/H$, (b) resistivity and (c) Hall mobility in LuB$_{12}$ (#3) at T=4.2 K.

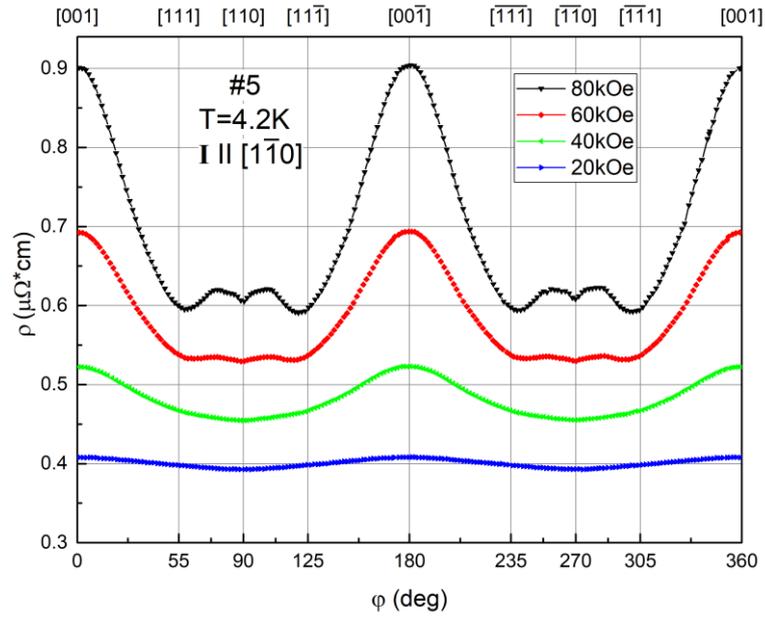

**Fig.S4.** Liquid helium angular dependences of resistivity in the LuB$_{12}$ (sample #5, **I**//[1-10]).

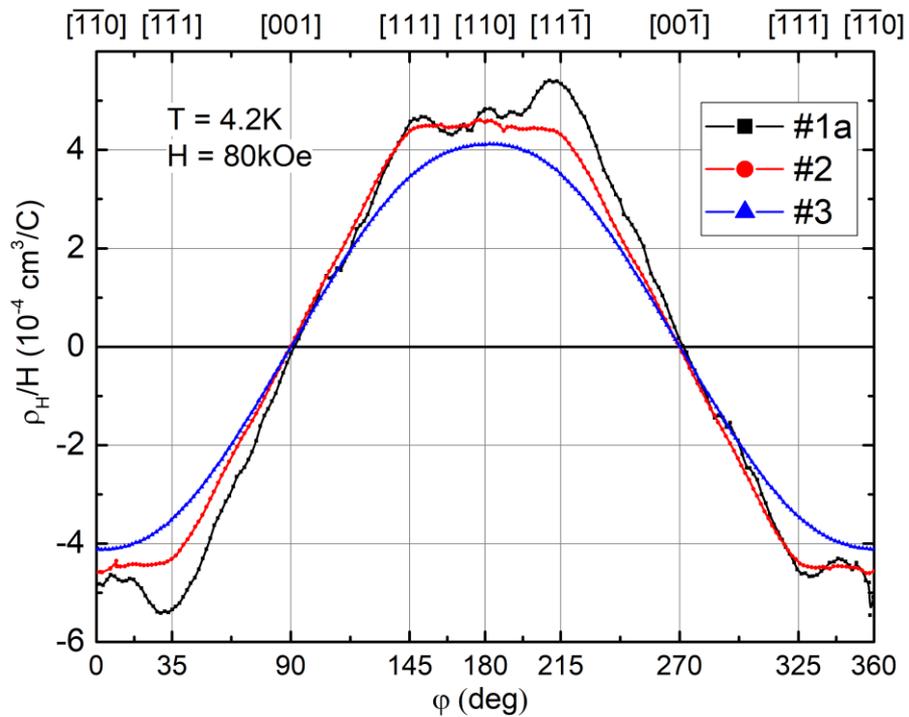

**Fig.S5.** The comparison of the reduced Hall resistivity angular dependences for the samples #1, #2 and #3 of LuB$_{12}$ (T=4.2 K, H= 80 kOe, **I**//[1-10], **n**//[110]).

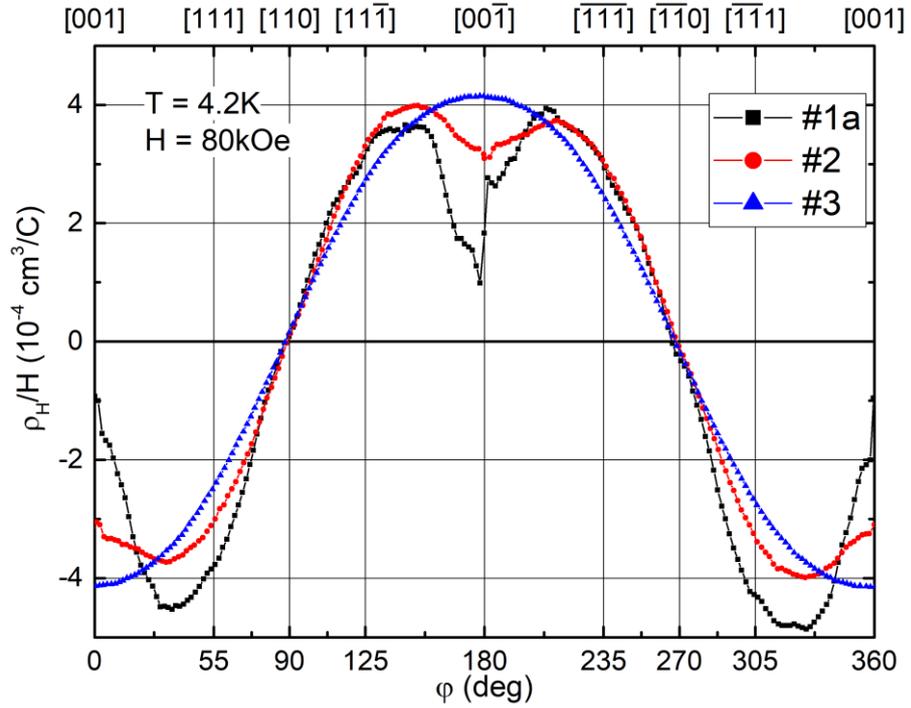

**Fig.S6.** The comparison of the reduced Hall resistivity angular dependences for the samples #1, #2 and #3 of LuB$_{12}$ (T=4.2 K, H= 80 kOe, **I**//[1-10]).

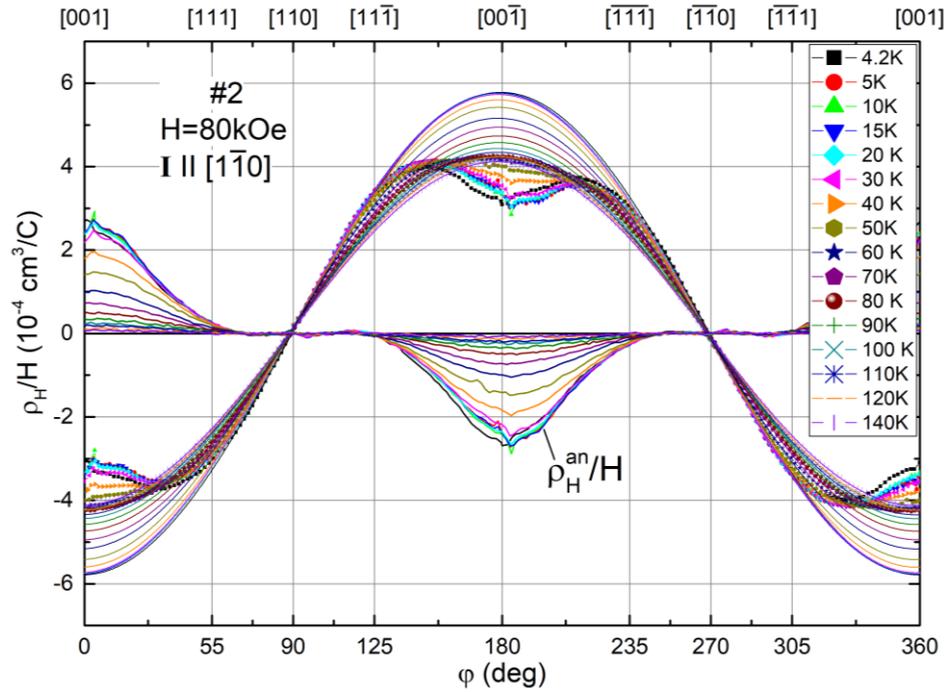

**Fig.S7.** The separation of the ordinary (cosin-type) and anomalous components in the reduced Hall resistivity at temperatures in the range 4.2-140 K at H=80 kOe for the sample #2 of LuB$_{12}$ (**I**//[1-10]).

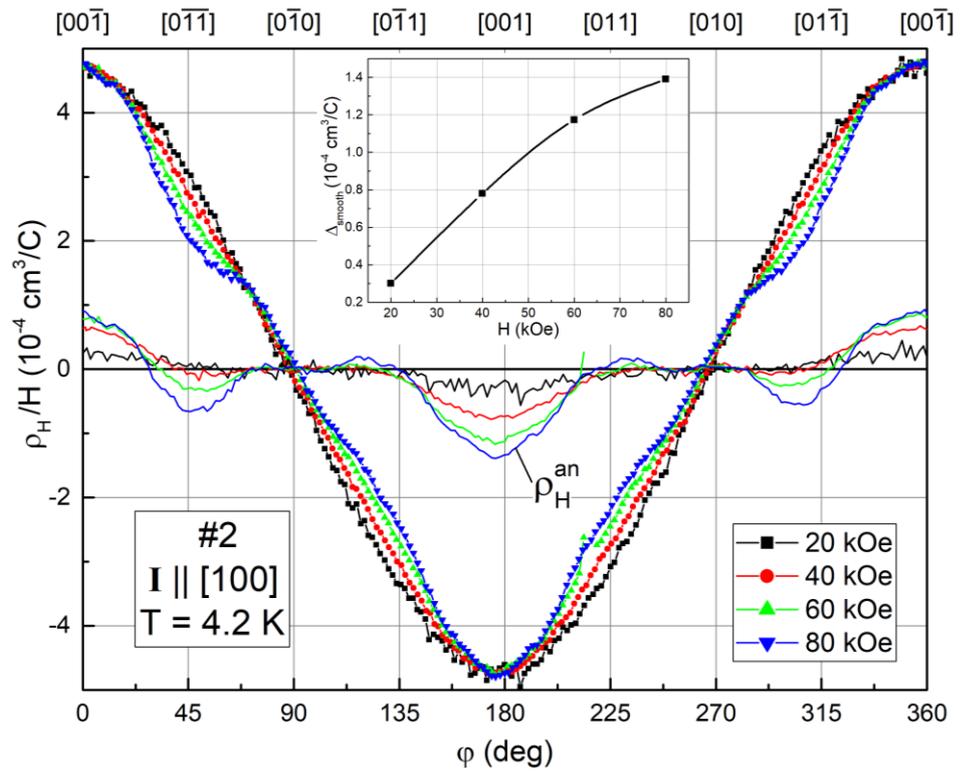

**Fig.S8.** The separation of the ordinary (cosin-type) and anomalous components in the reduced Hall resistivity at T= 4.2 K and in the range H≤80 kOe for the sample #2 of LuB$_{12}$ (**I**//[100]). Inset shows the amplitude of the anomalous contribution $\Delta_{smooth}$(H).

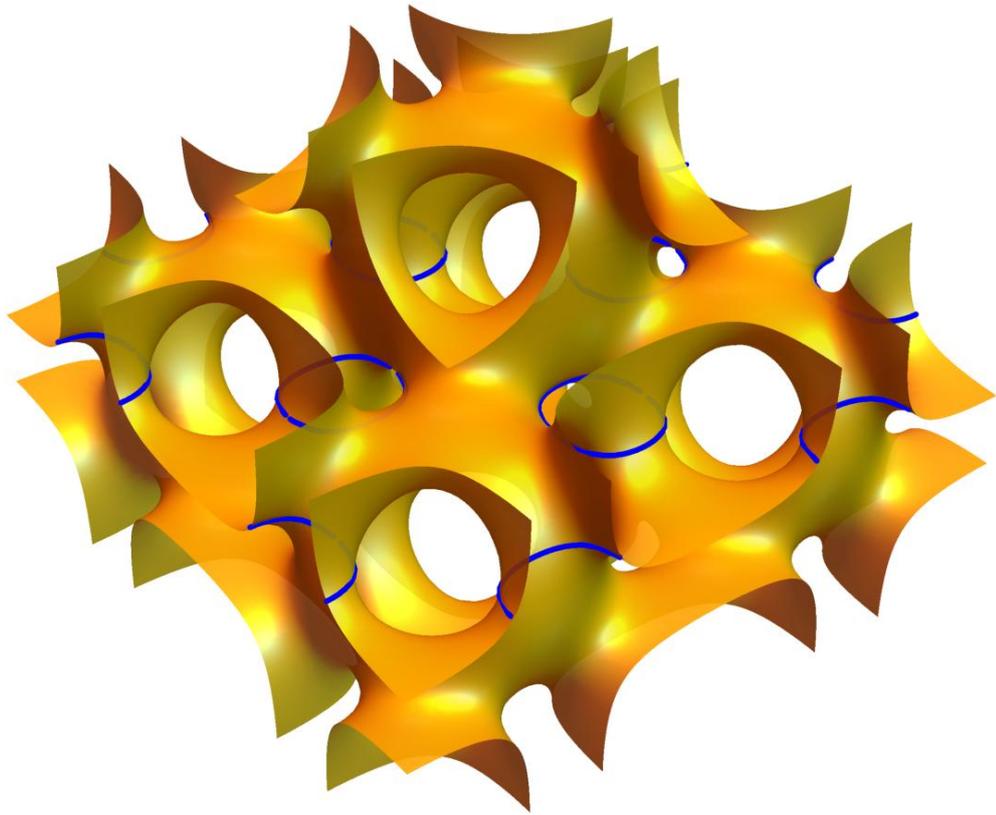

**Fig.S9.** The hole-like Fermi surface sheet in LuB$_{12}$. Solid blue lines show the four cornered rosettes (R$_{001}$ orbit (SdH $\alpha_2$ branch), see text for more detail).

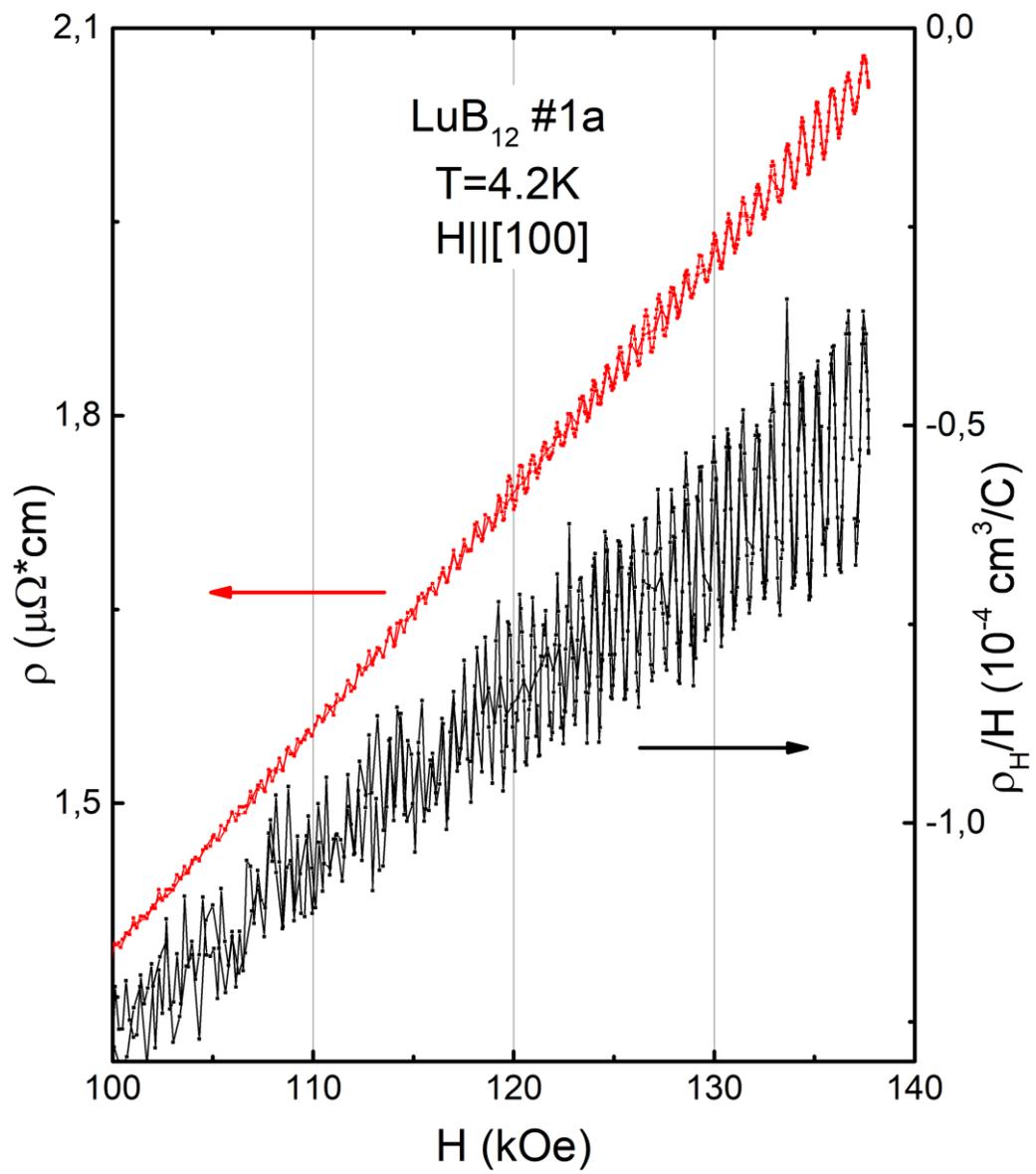

**Fig.S10.** Quantum oscillations of resistivity and the reduced Hall resistivity in LuB$_{12}$ (sample #1a).